\documentclass[apj]{emulateapj-rtx4}
\usepackage{natbib}
\usepackage{multirow}
\usepackage{fixltx2e}
\usepackage{float}



\begin{document}

\title{Complex {\rm Lyman}-$\alpha$ Profiles in Redshift 6.6 ultraluminous Lyman Alpha Emitters\altaffilmark{1,2}}

\author{
A.~Songaila\altaffilmark{3}, 
E.~M.~Hu\altaffilmark{3},
A.~J.~Barger\altaffilmark{4,5,3}, 
L.~L.~Cowie\altaffilmark{3},
G.~Hasinger\altaffilmark{3},
B.~Rosenwasser\altaffilmark{4},
C.~Waters\altaffilmark{3}
}

\lefthead{Songaila et al.}

\altaffiltext{1} {Based on data collected at the Subaru Telescope, which is operated
by the National Astronomical Observatory of Japan.}
\altaffiltext{2}{The W.~M.~Keck Observatory is operated as a scientific
partnership among  the California Institute of Technology, the University
of California, and NASA, and was made possible by the generous financial
support of the W.~M.~Keck Foundation.}
\altaffiltext{3}{Institute for Astronomy, University of Hawaii,
2680 Woodlawn Drive, Honolulu, HI 96822.}
\altaffiltext{4}{Department of Astronomy, University of Wisconsin-Madison,
475 N. Charter Street, Madison, WI 53706.}
\altaffiltext{5}{Department of Physics and Astronomy, University of Hawaii,
2505 Correa Road, Honolulu, HI 96822.}

\slugcomment{Accepted by The Astrophysical Journal}


\begin{abstract}
We report on a search for ultraluminous Ly$\alpha$ emitting galaxies (LAEs) 
at $z=6.6$ using the NB921 filter on Hyper Suprime-Cam on the Subaru telescope. 
We searched a $30~{\rm deg}^2$ area around the North Ecliptic Pole, which we observed 
in broadband $g', r',  i', z',$ and $ y'$ and narrowband NB816 and NB921, for sources 
with NB921 $<23.5$ and $z'$-NB921 $>1.3$. This corresponds to a selection of 
$\log L$(Ly$\alpha)>43.5$~erg~s$^{-1}$. We followed up seven candidate LAEs 
(out of thirteen) with the Keck DEIMOS spectrograph and confirmed five 
$z=6.6$ LAEs, one $z=6.6$ AGN with a broad Ly$\alpha$ line and a strong red 
continuum, and one low-redshift ([\ion{O}{3}]5007) galaxy. 
The five ultraluminous LAEs have wider line profiles
than lower luminosity LAEs, and one source, NEPLA4, has a complex line profile
similar to that of COLA1. 
In combination with previous results, we show that the 
line profiles of the $z=6.6$ ultraluminous LAEs are systematically different
from those of lower luminosity LAEs at this redshift. This result
suggests that ultraluminous LAEs generate 
highly ionized regions of the intergalactic medium in their vicinity that
allow the full Ly$\alpha$ profile of the galaxy---including any
blue wings---to be visible. If this interpretation is correct, then ultraluminous LAEs 
offer a unique opportunity to determine the properties of the ionized zones
around them, which will help in understanding the ionization of the $z \sim 7$ 
intergalactic medium.
A simple calculation gives a very rough estimate of 0.015 for the escape fraction of ionizing
photons, but more sophisticated calculations are needed to fully characterize 
the uncertainties.
\end{abstract}

\keywords{cosmology: observations 
--- galaxies: distances and redshifts --- galaxies: evolution
--- galaxies: starburst}


\section{Introduction}
\label{secintro}

Enormous progress is being made in finding and characterizing galaxies at very high redshift 
\citep[e.g.,][]{mclure13,bouwens15,finkelstein15}.
As well as their intrinsic interest, and the light they can shed on galaxy evolution, such high-redshift 
galaxy populations have the potential to allow the exploration of one of the key stages of cosmic history, 
the epoch of reionization (EoR) when the intergalactic gas (IGM) transitioned from being neutral to 
being very highly ionized.  In particular, since Ly$\alpha$ emission emerging from galaxies can be 
modified by the radiative damping wings of a neutral IGM, the presence of Ly$\alpha$ emitting 
galaxies (LAEs) at $z > 6$ can potentially pinpoint  when reionization occurred 
\citep[e.g.,][]{fontana10,stark10}.

The neutral hydrogen fraction in the surrounding IGM affects both the  strength and the shape 
of Ly$\alpha$ emission from distant galaxies \citep{haiman02, haiman05}.  For the purposes 
of identifying when reionization occurred, the Ly$\alpha$ luminosity function (LF)
itself is not enough, since both it and the UV continuum LF will
change as the star formation rate density (SFRD) evolves.  However, the LAE
fraction---the number of LAEs relative to
the number of continuum-selected galaxies---is invariant to the SFRD
evolution and has the potential to be a strong indicator of the reionization 
boundary \citep[e.g.,][]{robertson10}, 
as do the properties of the individual Ly$\alpha$ lines.

However, it is also possible that any evolution of the LAE fraction
may be driven by changes in the intrinsic properties of the galaxy
population rather than by an increasing 
neutrality of the IGM \citep[e.g.,][]{stark15,sobral17}. Modeling shows that 
outflows lead to a redshifted Ly$\alpha$ 
emission line profile, while inflows produce a blueshifted profile, which will 
be preferentially scattered by the IGM \citep[e.g.,][]{neufeld90}.

Thus, the gas kinematics in these high-redshift galaxies is a key element
in determining how the Ly$\alpha$ emission interacts with the IGM
and what the final Ly$\alpha$ profile will look like. 
Most high-redshift LAEs show an asymmetric profile with a red tail
and a fast cutoff to the blue, but this asymmetry could  
reflect either the output shape from the galaxy, or the
IGM cutting off the blue side of an initially more symmetric
profile. In addition, the width of the line depends on the inflow and 
outflow velocities,
and changes in the width will modify the amount of IGM scattering.
Evolution in the galaxy properties can therefore reproduce the observed 
LAE evolution with a much smaller change in the neutral fraction in the IGM 
\citep[e.g.,][]{stark15}.

Substantial samples of LAEs out to $z=6.6$ have now been found using 
narrowband searches 
\citep[e.g.,][]{hu10,ouchi10,matthee15,santos16,bagley16,konno17,zheng17},
and a handful of sources have been detected at $z\sim7.3$ \citep{konno14}.
The highest redshift securely detected LAE is at $z=8.683$ \citep{zitrin15}.
All groups find the Ly$\alpha$ LF at lower luminosities ($\log L({\rm Ly}{\alpha})<43.5$) 
to be declining with increasing redshift, and it may be dropping very rapidly at $z=7$
\citep{zheng17}.
There is also evidence of a drop in the LAE fraction beyond $z=7$ that has been 
interpreted as evidence for a rapid change in the IGM neutral hydrogen fraction 
near this redshift \citep[e.g.,][]{stark10,konno14}.
At these lower luminosities,
the shapes and velocity widths of the Ly$\alpha$ lines are all extremely similar
and show little evolution with redshift between $z=5.7$ and $z=6.6$ \citep{hu10, matthee17b}.

However, recent work has suggested that the behavior of the very 
highest luminosity 
LAEs may be different. First, their LF may show little evolution over 
the $z=5-7$ range \citep{santos16}.
Second, \citet{stark17} have found Ly$\alpha$ emission in 
all of a sample of four luminous galaxies at $z>7$ selected on the
basis of their IRAC colors 
\citep[which indicates probable nebular emission; ][]{roberts-borsani16}.
Thus the most luminous galaxies may have less evolution in
their LAE fraction than the lower luminosity samples. 

What is even more intriguing is that the most luminous LAE---COLA1 at $z=6.6$ with
$\log L({\rm Ly}{\alpha})\sim44$~erg~s$^{-1}$ \citep{hu16}---is the first source
near the EoR  to show a complex LAE profile with a 
stronger red and weaker blue component
(though there has been debate over whether it might be
a lower redshift emission line; \citealt{matthee17b}).
We shall loosely refer to such sources as blue-wing LAEs;
however, without an accurate measure of the systemic velocities,
we do not know if the blue component is at a negative velocity
with respect to the galaxy.
These data may suggest that the \ion{H}{2} regions 
being generated around the highest luminosity LAEs are ionized 
enough to permit the transmission of Ly$\alpha$,
allowing the LAE to be visible even as the IGM 
as a whole becomes more fully neutral \citep[e.g.,][]{matthee15}.
If this is correct, then we may be able to estimate the escaping ionizing photon
production in the parent galaxy from the required properties of the \ion{H}{2} 
region.
The key to determining this is to obtain large samples of the most luminous
LAEs and to measure spectroscopically their Ly$\alpha$ line properties as 
a function of luminosity. 

This is one of the goals of our HEROES survey, which 
aims to survey a $120~{\rm deg}^2$ area around the North Ecliptic Pole (NEP).  
HEROES will provide a large homogeneous galaxy sample with which to 
determine the properties of LAEs at $z=5.7$ and $z = 6.6$ over a volume large 
enough to enable us to find significant samples of the most luminous
LAEs ($\log L({\rm Ly}{\alpha})\sim43.5-44$~erg~s$^{-1}$) and to determine the 
Ly$\alpha$ line properties using follow-up spectroscopy. This search requires a 
very large area survey but only a relatively modest depth in flux sensitivity.

In the present paper, we describe our candidate selection of $z=6.6$ LAEs with 
NB921 brighter than 23.5 (AB) in the first $30~{\rm deg}^2$ region of HEROES. 
Out of our 13 candidates, we have followed up spectroscopically the 7 with $z'$-band detections (since we are secure that these are not spurious) and have confirmed most of them as $z = 6.6$ LAEs.
We find that these ultra-luminous LAEs have wider and more complex 
profiles than the lower luminosity $z=6.6$ LAEs. We have also discovered a 
second LAE with a blue wing profile similar to COLA1.

We assume $\Omega_M=0.3$, $\Omega_\Lambda=0.7$, and
$H_0=70$~km~s$^{-1}$~Mpc$^{-1}$ throughout.
All magnitudes are given in the AB magnitude system,
where an AB magnitude is defined by
$m_{AB}=-2.5\log f_\nu - 48.60$.
Here $f_\nu$ is the flux of the source in units of
erg~cm$^{-2}$~s$^{-1}$~Hz$^{-1}$.

\section{Data}
\label{secdata}

The images used in this work were obtained as part of our HEROES survey 
(Hasinger et al.\ 2018, in preparation), which uses the wide-field capability 
($1.5^\circ$ diameter field-of-view) of the Hyper Suprime-Cam camera 
\citep[HSC;][]{miyazaki12} on the Subaru 8.2~m telescope to map an area of 
120~${\rm deg}^2$ around the NEP. The survey was designed  to match the deepest 
X-ray observations that will be obtained with the eROSITA mission \citep{merloni12}.
HEROES includes broadband $g', r', i', z',$ and $y'$ images and narrowband 
NB816 and NB921 images with HSC and $U$ and $J$ images with
MegaPrime/MegaCam and WIRCam on the
3.6~m Canada-France-Hawaii Telescope (CFHT).  The $r'$ and $i'$ data were taken with the HSC-r2 and HSC-i2 filters, installed in 2016.

\begin{figure}[ht]
\centerline{\includegraphics[width=9cm,angle=0]{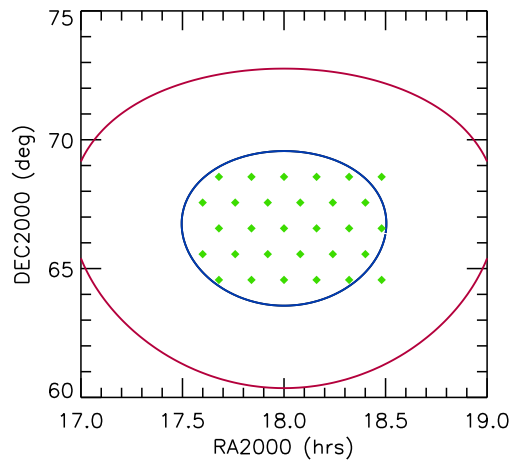}}
\caption{The close-packed stepping pattern of $1^{\circ}$ offsets adopted for 
the $6 \times 5$ point grid covered in these NEP observations, with central 
position R.A.: 18.000~hr, Decl.: 66.5607$^\circ$. 
The blue circle shows the area of the current observations, and the red circle
the area of the full HEROES survey. 
\label{pattern}
}
\end{figure}

\begin{figure}[ht]
\centerline{\includegraphics[width=7cm,angle=0]{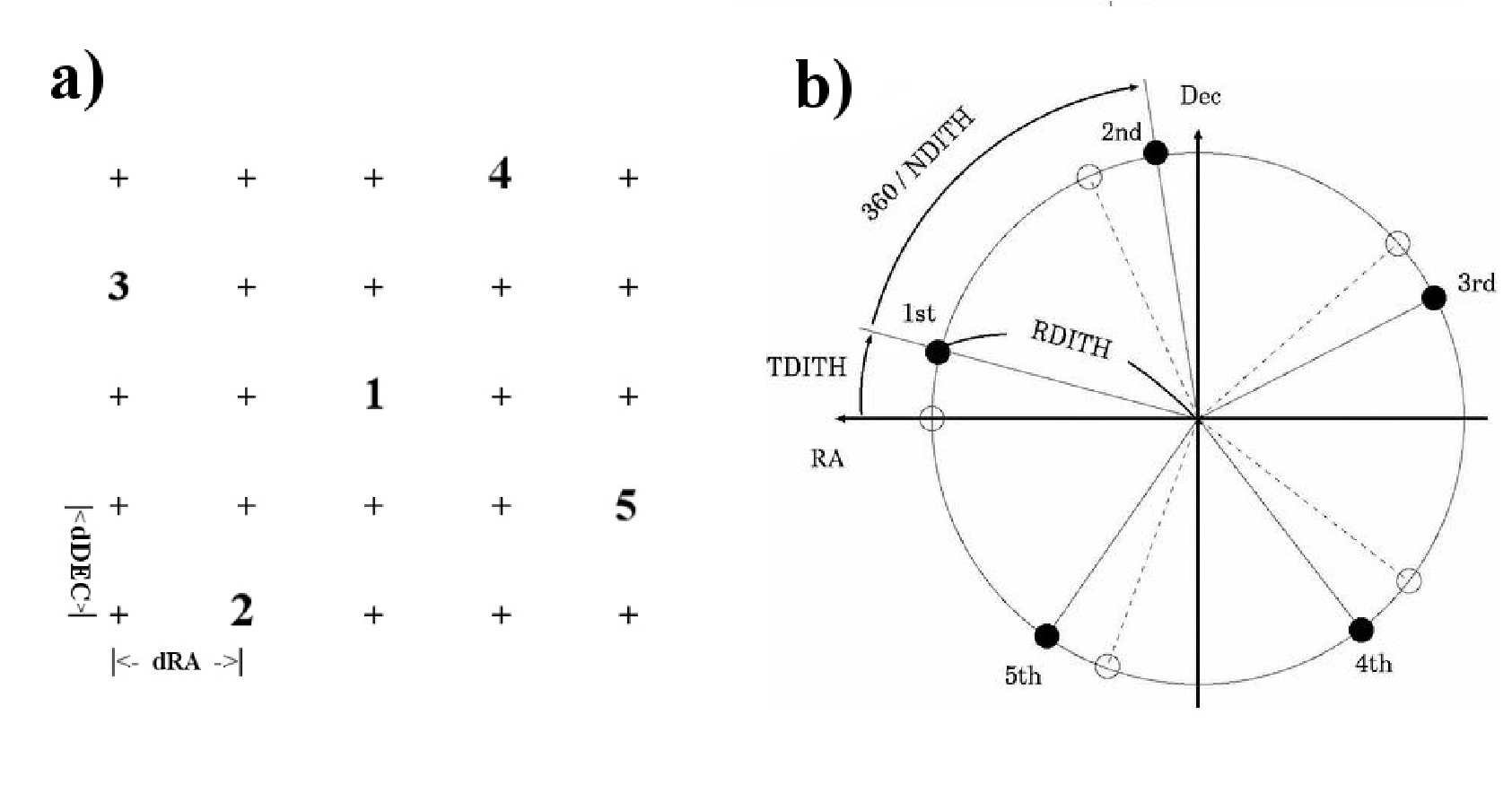}}
\caption{Dither patterns from the HSC queue mode Program PI Document. 
We adopted the N-Shot pattern (panel (b)) with NDITH=5 for our HEROES observations.
\label{dither}
}
\end{figure}

The HSC observations of the NEP field were taken in a close packed
set of observation stepped at 1~deg offsets. 
We show the stepping pattern in Figure~\ref{pattern}.
At each position we observed using a 5-point 
mosaic pattern in N-shot mode (see Figure~\ref{dither}).  The 5 positions 
spaced equally around a circle of radius $120\arcsec$ (RDITH =120) 
from the field center, with the first position rotated 
clockwise by $15^{\circ}$ (TDITH = 15) from the 
east-west line (R.A. offset $120\arcsec$ east of the center).  

Given the $1.5^\circ$ diameter field-of-view of HSC, 
the combination of our stepping pattern of staggered  
$1^{\circ}$ offsets between field centers and the N-Shot mosaic 
pattern with NDITH = 5 gives slightly more uniform field coverage 
over the wide NEP region than any other possible pattern.
This approach ensures that every point in the center of the $6 \times 5$ grid 
of the currently observed NEP field is also covered by the 
mosaic observations at four adjacent pointings centered 
at corners diagonally $1.41^{\circ}$ away from the center pointing, 
and that as the outer regions of the sample field get expanded
with new observations,  this
also holds true for the extended field.

The HSC observations with the NB921 filter were taken on UT dates 
2016-08-02, 2016-08-03, and 2016-08-05, and again on 2017-06-22. 
During 2016, on the first night, when the first 14 positions were taken,  
the median seeing in NB921 was $\sim 0\farcs85$, and on 
the second night, $0\farcs6$ or better.  On the third night, there was  
heavy extinction (4~mags) with median seeing of $\sim1\farcs1$.
As a result, this  part of the field was re-observed in NB921 on UT 2017-06-22,
which was a photometric night with median seeing of $\sim0\farcs54$.

Matching $z'$ observations were obtained on UT dates 2016-07-06, 
2016-07-11, 2016-08-05, and 2017-06-28, with median seeing
of $\sim 0\farcs58$, $\sim 0\farcs64$, $\sim1\farcs1$, and $\sim0\farcs58$, 
respectively.

All  HSC images were reduced with the Pan-STARRS Image Processing Pipeline \citep[IPP,][]{magnier2017.datasystem}. 
We chose to use this rather than the HSC pipeline because it is already set
up and tested on our computer system. However, both rely on the Pan-STARRS
data for astrometric and photometric calibration and we expect that
there will be little difference between the final calibrated images.
The IPP pipeline is well tested, and, with only minor configuration 
additions, can operate on any imaging data set. In addition, the availability of a dedicated 
computer cluster allows the large data volume to be processed quickly. The details of pixel 
processing are presented in \citet{waters2017}. Briefly, each exposure has the instrument 
signal removed, and then photometry is performed on the individual exposures. The objects 
detected are matched against the Pan-STARRS reference catalog to determine the astrometric 
and photometric calibration for the exposure.

\begin{figure}[H]
\centerline{\includegraphics[width=10cm,angle=0]{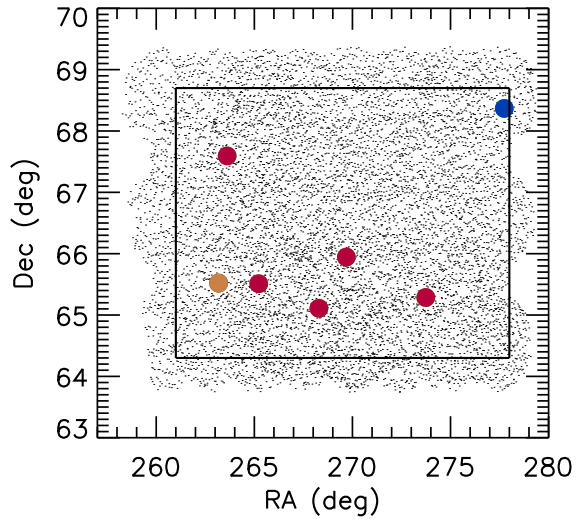}}
\caption{The shaded area shows the full $45~{\rm deg}^2$\ field
of the current NEP observations. The solid rectangle shows the more uniform 
$30~{\rm deg}^2$\ area used in the present paper.
Filled circles show the five spectroscopically confirmed LAEs (red), the one 
high-redshift AGN (gold),
and the one [\ion{O}{3}] emitter (blue). 
See Table~\ref{tabobj} for details.
\label{show_objects}
}
\end{figure}

From these solutions, individual exposures that have good seeing and transparency are 
transformed to a common pixel grid and co-added into stacked frames.  For this analysis, 
the seeing limit excluded any input with PSF FWHM greater than $1\farcs36$, and the 
transparency limit excluded any input with a measured zeropoint 0.3~mag brighter than 
the median. The stacking process removes non-astronomical artifacts, such as reflection 
glints, while enhancing the signal to noise (S/N) of faint sources. These faint sources are 
measured by additional stack photometry run across all filters simultaneously 
\citep{magnier2017.analysis}.  Measurements are forced in all filters if a $5\sigma$ 
significant source is detected in any single band.

Only the central observations covering 45~${\rm deg}^2$ have been obtained so far 
(Figures~\ref{pattern} and  \ref{show_objects}). The catalog of objects 
detected with  S/N $>5\sigma$ in any band throughout this 
area contains 23.9 million objects.
From this catalog, we selected 2.8 million objects with NB921 Kron 
magnitudes brighter than 23.5 lying in the most uniformly covered 
area of the field (${\rm R.A.}=261^{\circ} - 278^{\circ}$, 
${\rm Decl}=64.3^{\circ} - 68.7^{\circ}$), shown with the
rectangle in Figure~\ref{show_objects}. The final sample used in this paper
was taken from the most uniform area of 30~${\rm deg}^2$.

\begin{figure}[H]
\centerline{\includegraphics[width=10cm,angle=0]{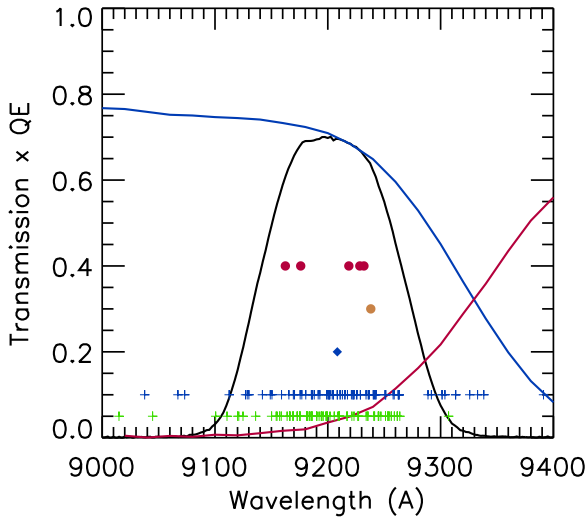}}
\caption{Product of filter transmission and CCD quantum efficiency for 
the HSC filters NB921 (black), $z'$ (blue), and $y'$ (red).  
Filled symbols show the five spectroscopically confirmed LAEs (red circles),
the one high-redshift AGN (gold circle), and the one [\ion{O}{3}] emitter 
(blue diamond). Crosses show spectroscopically confirmed random 
[\ion{O}{3}] (blue) and [\ion{O}{2}] (green) emitters in the field, all
shown at the position of the weighted mean of the doublet.
\label{921_filter}
}
\end{figure}

\begin{figure*}
\includegraphics[width=9.2cm,angle=0]{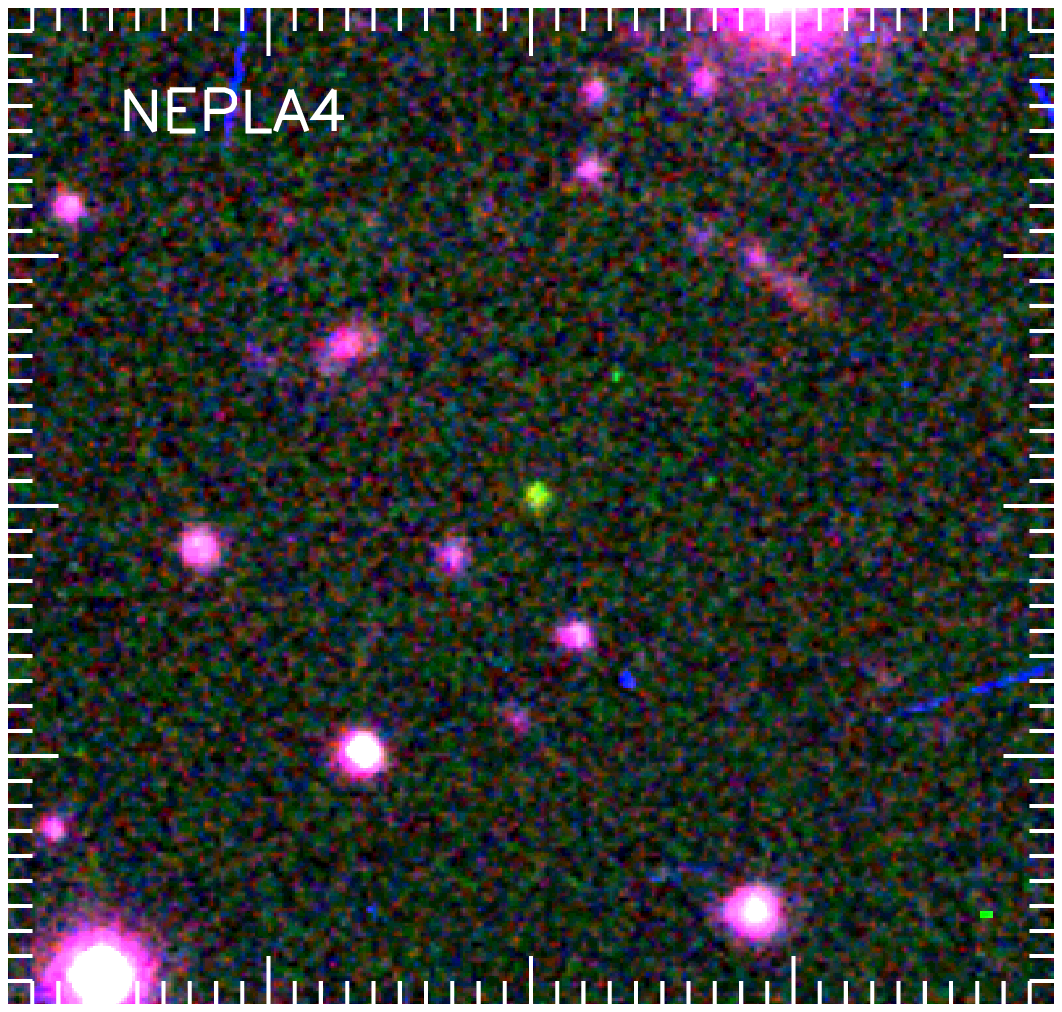}
\includegraphics[width=9.2cm,angle=0]{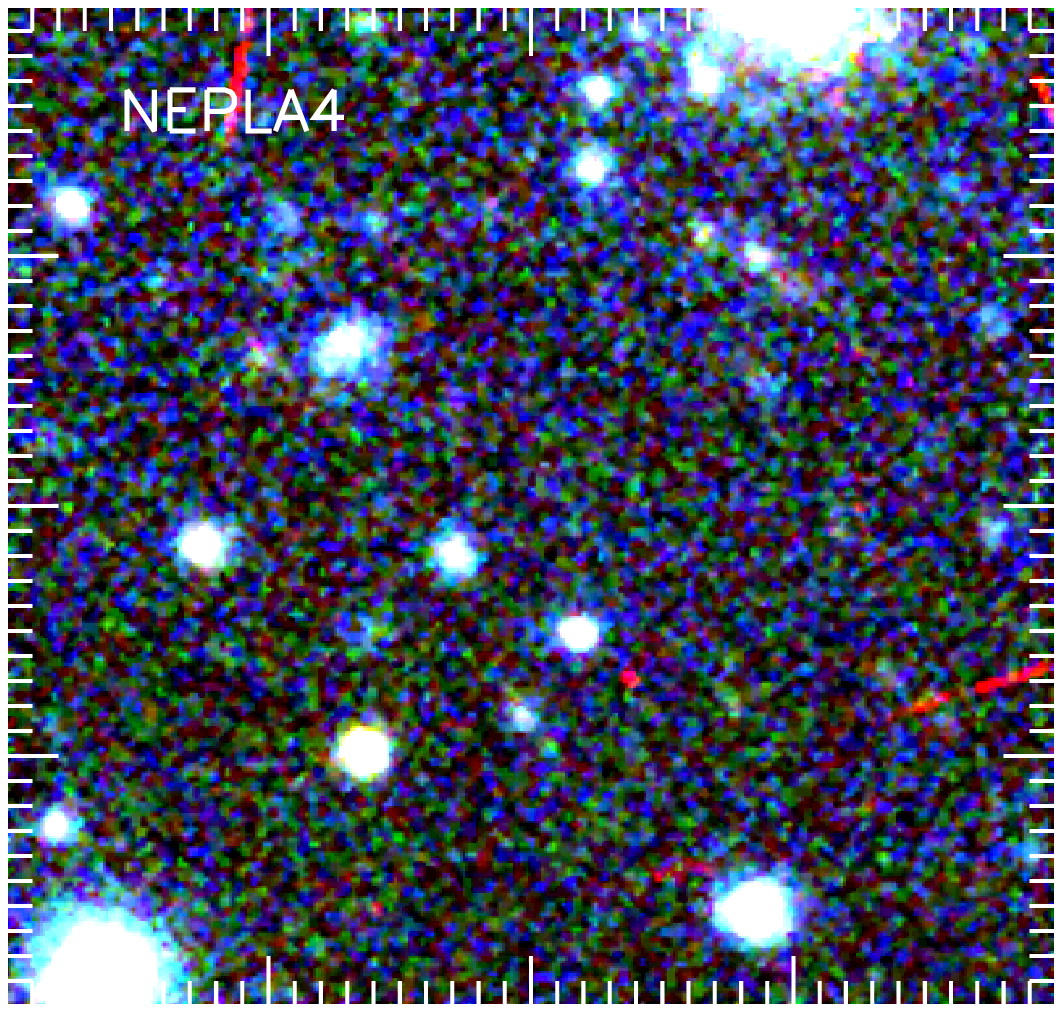}
\includegraphics[width=9.2cm,angle=0]{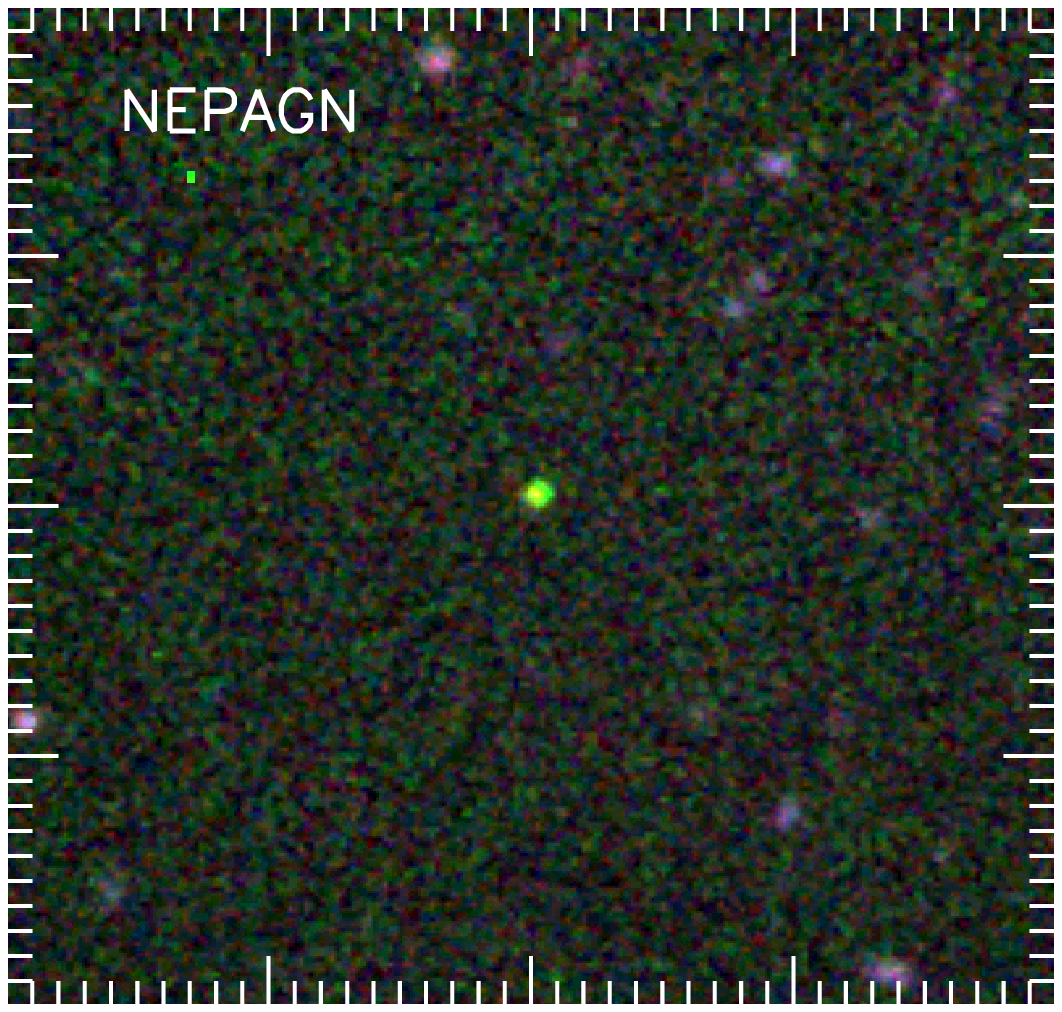}
\includegraphics[width=9.2cm,angle=0]{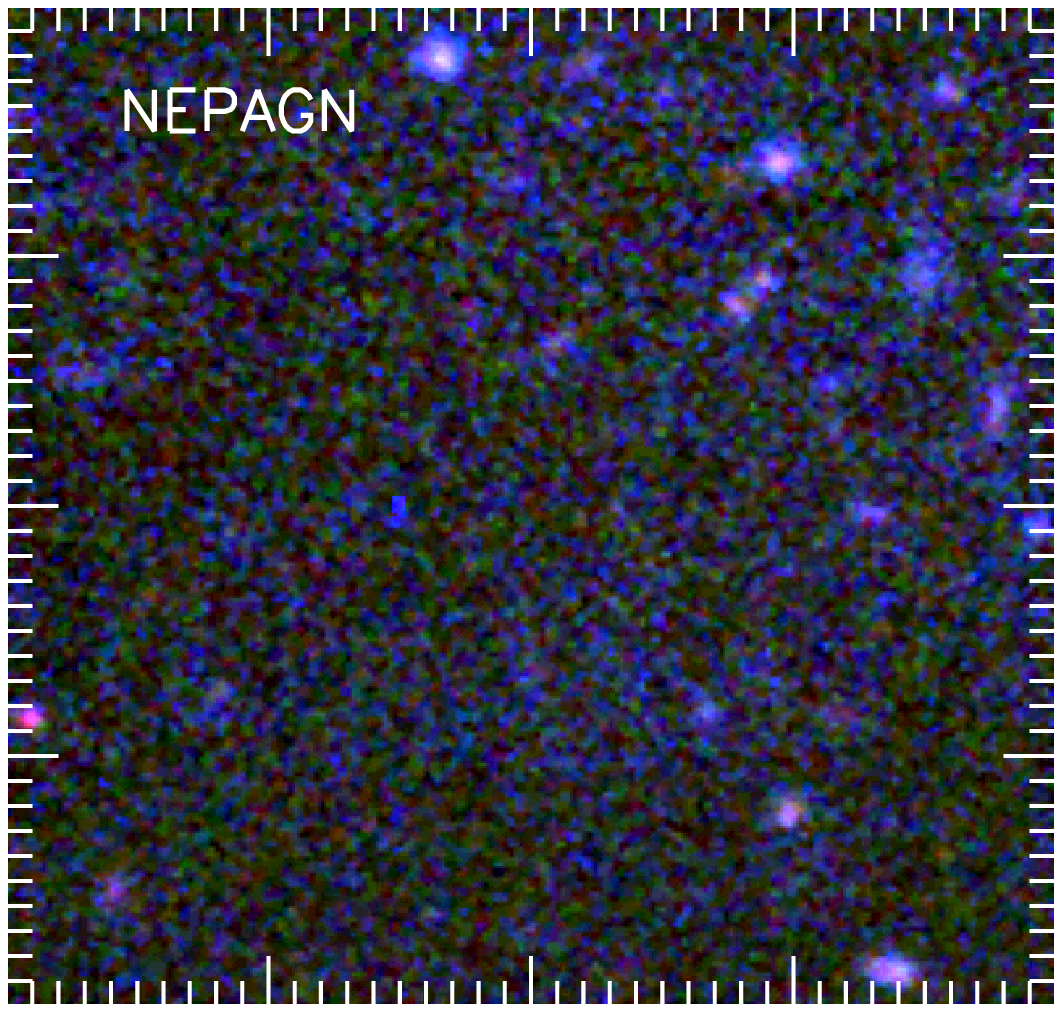}
\includegraphics[width=9.2cm,angle=0]{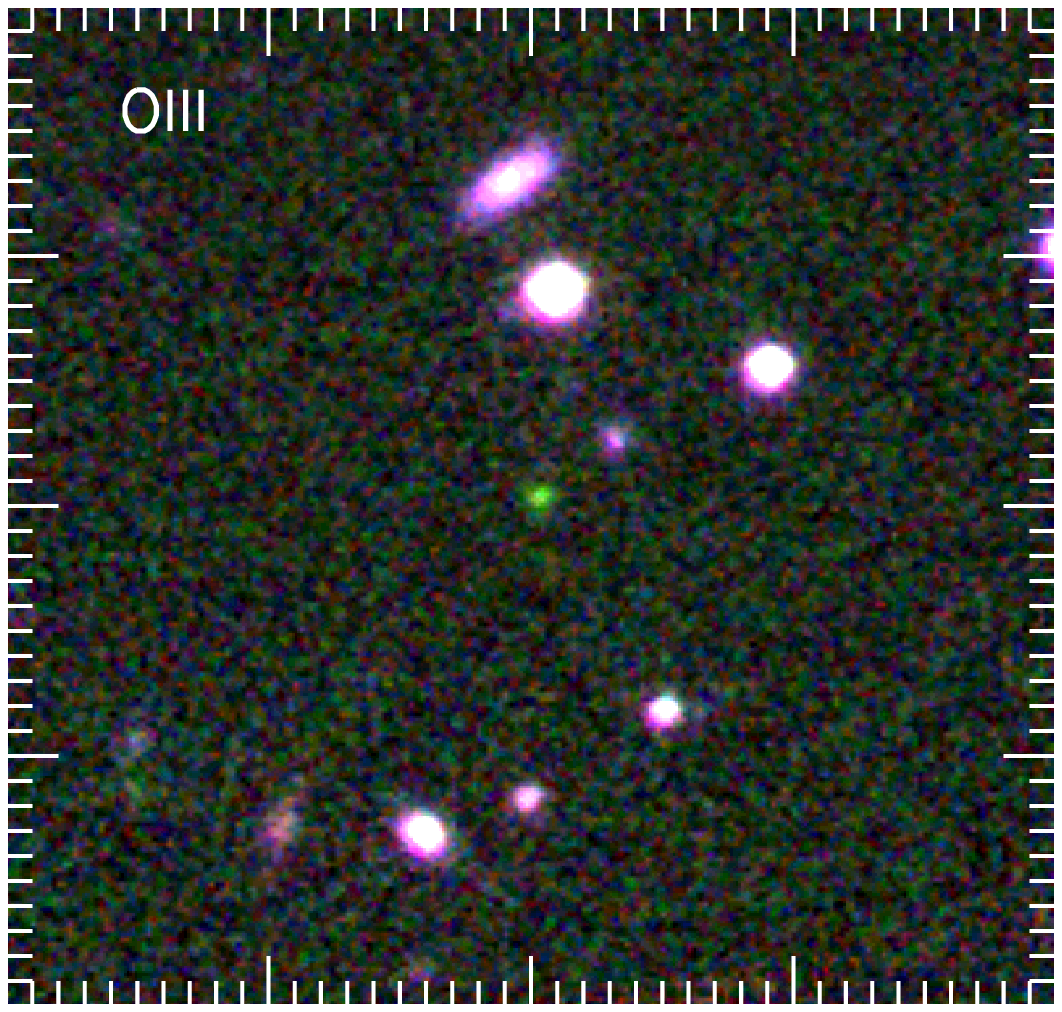}
\includegraphics[width=9.2cm,angle=0]{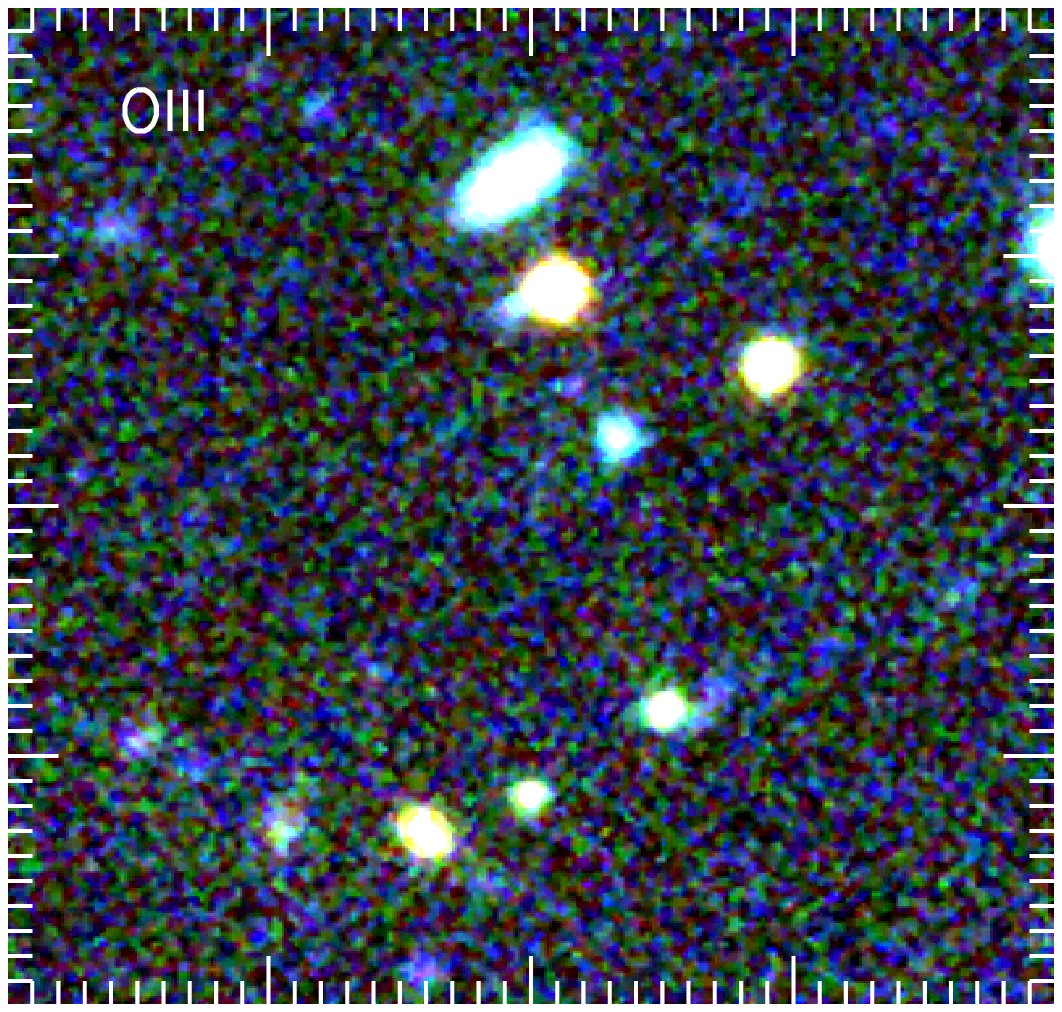}
\caption{{\it Left}:  HSC images of the confirmed LAE, NEPLA4 (top), 
the NEP AGN (middle), and an [\ion{O}{3}]5007 emitter (bottom). 
NB921 is shown in green, $z'$ in red, and $r'$ in blue. The
images are $30^{\prime\prime}$ on a side. 
{\it Right}:  Same sources shown in $g'$ (blue), $r'$ (green), and $i'$ (red).
}
\label{2d-images}
\end{figure*}

The exposure time in NB921 varies throughout the 30~${\rm deg}^2$ area, 
depending on the overlap and on edge effects. The mean exposure
per pixel is 5600~s with a range from 2200~s to 9000~s;
the lowest exposures are at the edges. The $1\sigma$ noise in a $2\arcsec$ 
diameter aperture ranges from 25.4 to 26.1. 
The adopted 23.5~mag selection in NB921 of LAE candidates provides a  
$>5\sigma$ criterion throughout the area and a much higher 
S/N (mean of $9\sigma$) through most of the area. The corresponding $z'$-band 
observations have exposure times from 2100~s to 9600~s (mean of 6200~s) 
and a $1\sigma$ noise of 26.2 to 27.0 (mean of 26.7 in a  $2\arcsec$ diameter 
aperture).

We next measured the magnitudes for all of the bands using $2\arcsec$
diameter apertures centered on the NB921 positions and
searched for $z=6.6$ LAE candidates with ($z'-$NB921)$>1.3$ 
that were not detected above a $2\sigma$ level
in any of the $g'$, $r'$  and $i'$ bands. The $z'$ 
filter bandpass is generally to the blue of NB921 (Figure~\ref{921_filter}),
but the red end covers NB921 and provides a
reasonable continuum measurement. The half maximum of the
transmission corresponds to a redshift range of $z=6.52-6.63$,
which we use in combination with the $30~{\rm deg}^2$ area
to measure the comoving volume of the survey ($2.7\times10^7$~Mpc$^3$).  
This is the largest volume yet surveyed for ultraluminous LAEs. It is
roughly 40\% larger than the SILVERRUSH survey of \cite{konno17}.

The LAE candidate sample chosen in this way still contains numerous artifacts 
from contamination by bright stars, moving objects, glints, etc. 
We next visually inspected the narrowband 
and broadband images of each of the LAE candidates and removed artifacts. At this stage, 
we also visually inspected a noise-weighted sum of the $g'$, $r'$, and $i'$ bands
and eliminated sources where we could see the narrowband source in 
the combination of these bluer bands. 

In the left panels of Figure~\ref{2d-images}, we show images of some of our candidates 
(NB921 in green, $z'$ in red, and $r'$ in blue). In the right panels, we show the same 
sources in $g'$ (blue), $r'$ (green), and $i'$ (red). These broadband data are much 
deeper than the narrowband data, and one can see that the candidates are absent 
in these images.

We prioritized the candidates that are also detected in the $z'$-band for 
spectroscopic follow-up, 
since we are secure that these are not spurious. This gives a sample
of seven primary sources, which we show in Figure~\ref{show_objects} 
and summarize in Table~\ref{tabobj}. There are a further
six possible candidates faint in $z'$. Since these sources
are only detected in NB921, they are more questionable and
may correspond to unidentified artifacts.  
We have not yet observed them spectroscopically
which would be necessary to confirm them.
However, 
they could, in principle, raise the LAE numbers to eleven vs.\ the current
five. We postpone a discussion of the bright end LAE LF until
we have completed the spectroscopic follow-up of the full
sample, but we note that, based on the photometric sample,
our results are roughly consistent with the \cite{konno17}
LF at $\log {L({\rm Ly}}\alpha) > 43.5~{\rm erg\ s}^{-1}$ but low
compared with that of \cite{santos16}.

\section{Spectroscopic follow-up}
\label{specs}

We obtained spectroscopy with DEIMOS on Keck~II of the seven primary targets
in Table~\ref{tabobj} during observing runs in June, August, October 2017 and March 2018.
The observations were made with the G830 grating using a $1{\arcsec}$
slit, giving a resolution measured from the sky lines of 83~km~s$^{-1}$ for the $z=6.6$ 
LAEs. Each exposure consisted of three 20~min sub-exposures dithered by $\pm 1{\farcs5}$ 
along the slit in order to obtain precise sky subtraction, which is critical given the
ghosting in this grating configuration. Total exposures
ranged from $1-3$~hr. The seeing was generally
$\sim 0{\farcs5}-0{\farcs8}$, and 
conditions were photometric, except for the observations
of NEPLA2 when there was variable transmission. 
The data were reduced with our standard pipeline \citep{cowie96}.
The three most luminous galaxies in the extended COSMOS
field - CR7 and MASOSA \citep{sobral15} and COLA1 \citep{hu16} - were observed in the same configuration,
allowing a detailed comparison of their spectra with the spectra of the present NEP sources.

\begin{figure*}
\includegraphics[width=8.0cm,angle=0]{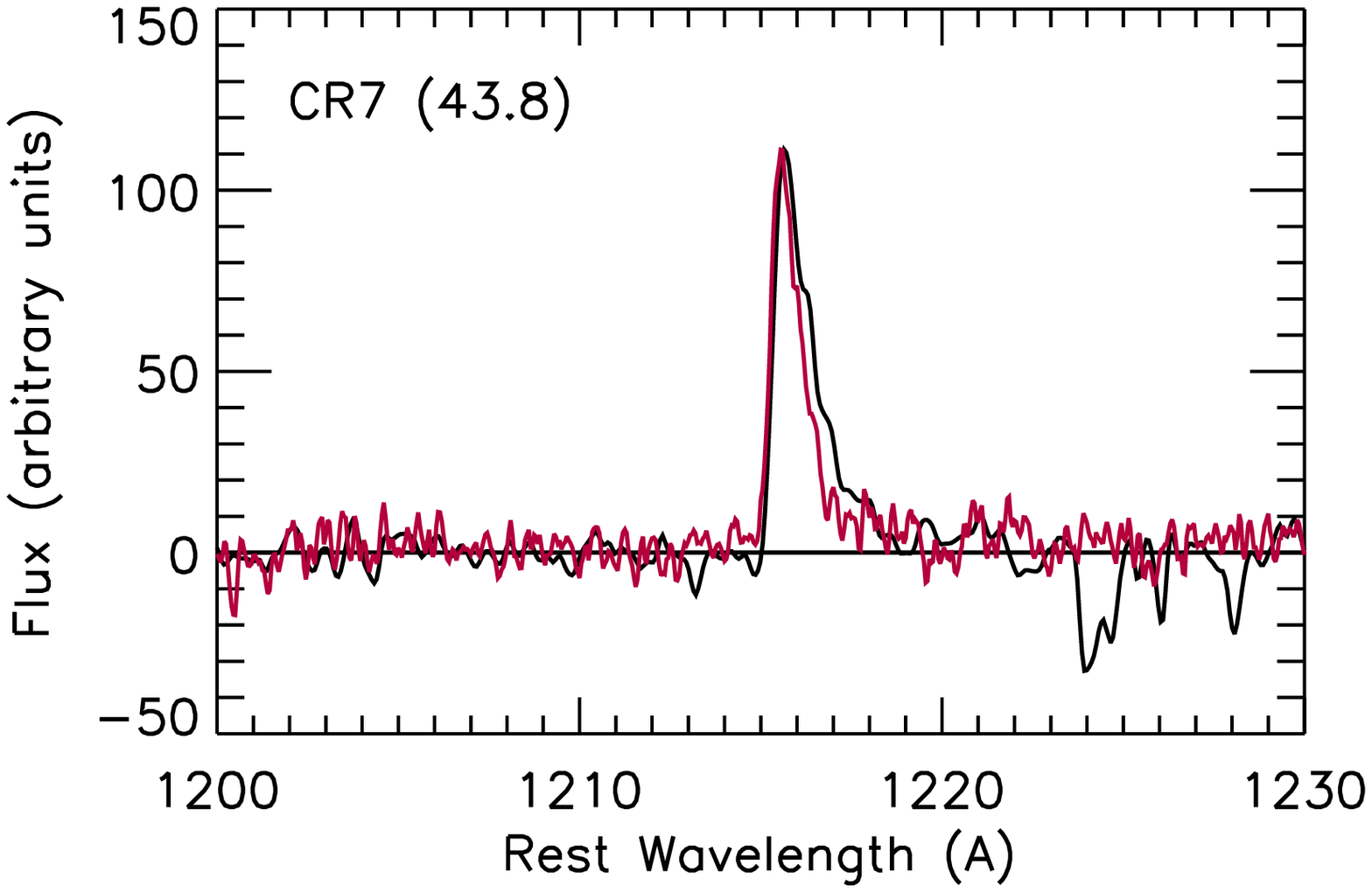}
\includegraphics[width=8.0cm,angle=0]{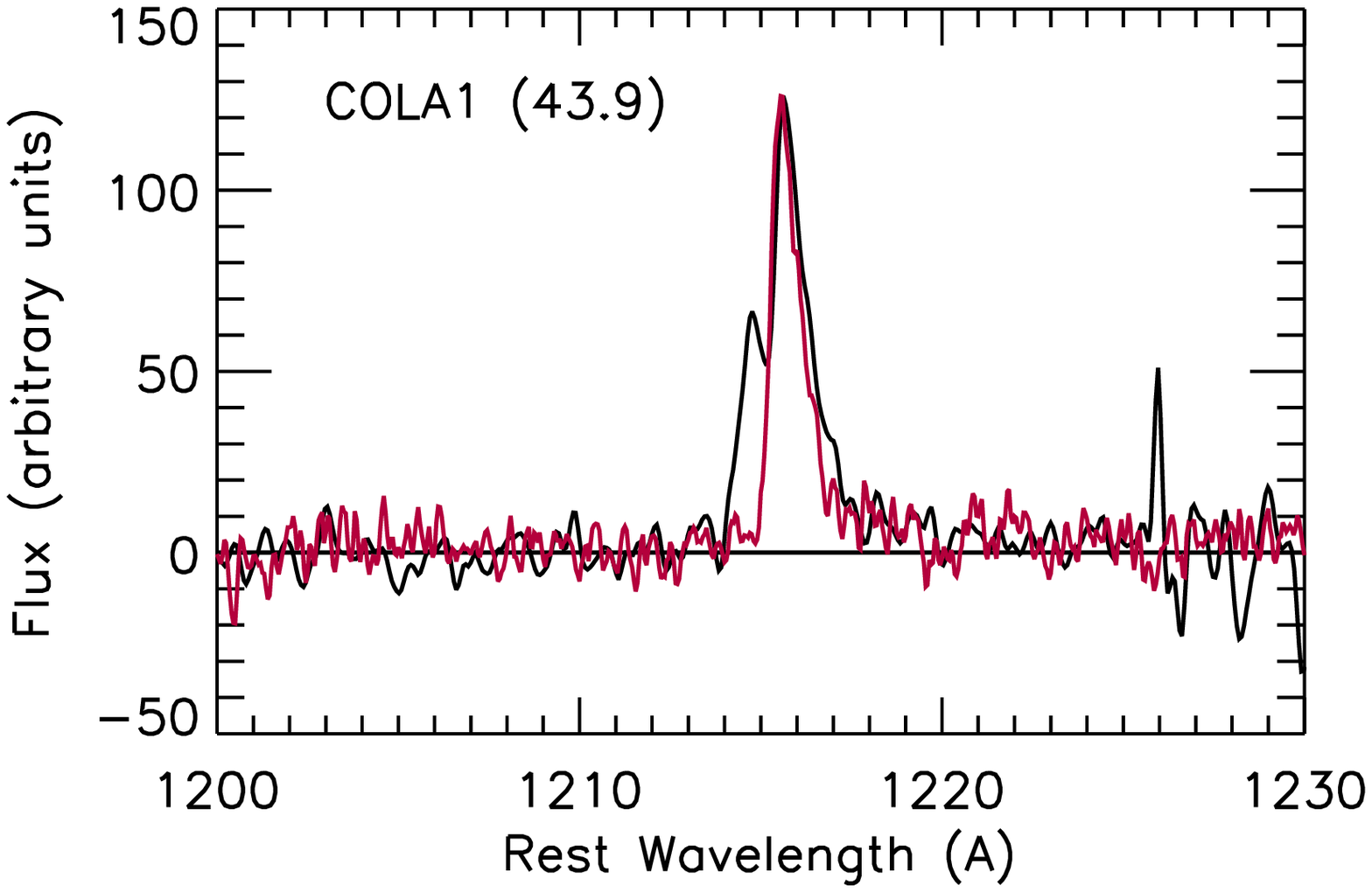}
\includegraphics[width=8.0cm,angle=0]{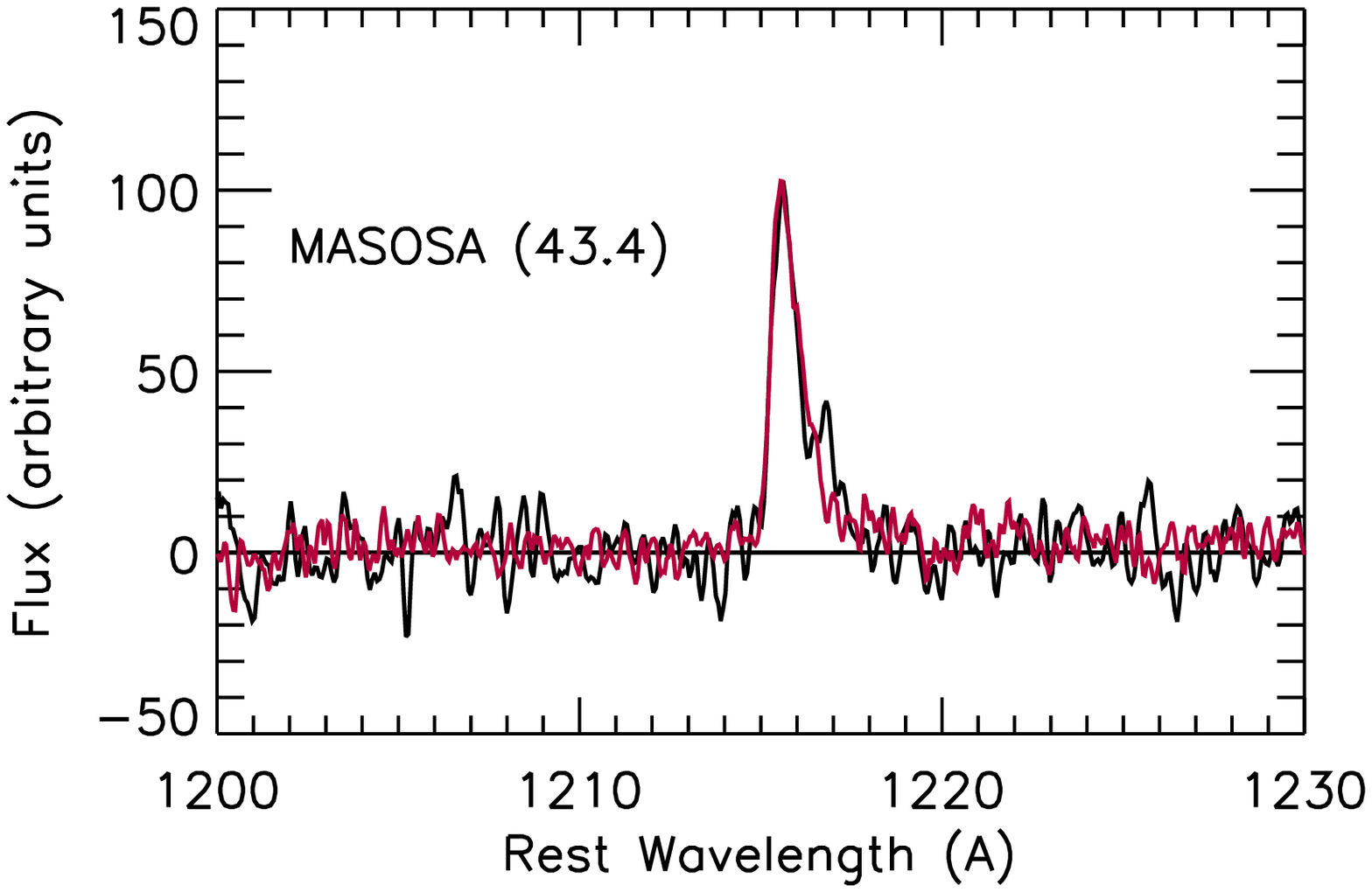}
\includegraphics[width=8.0cm,angle=0]{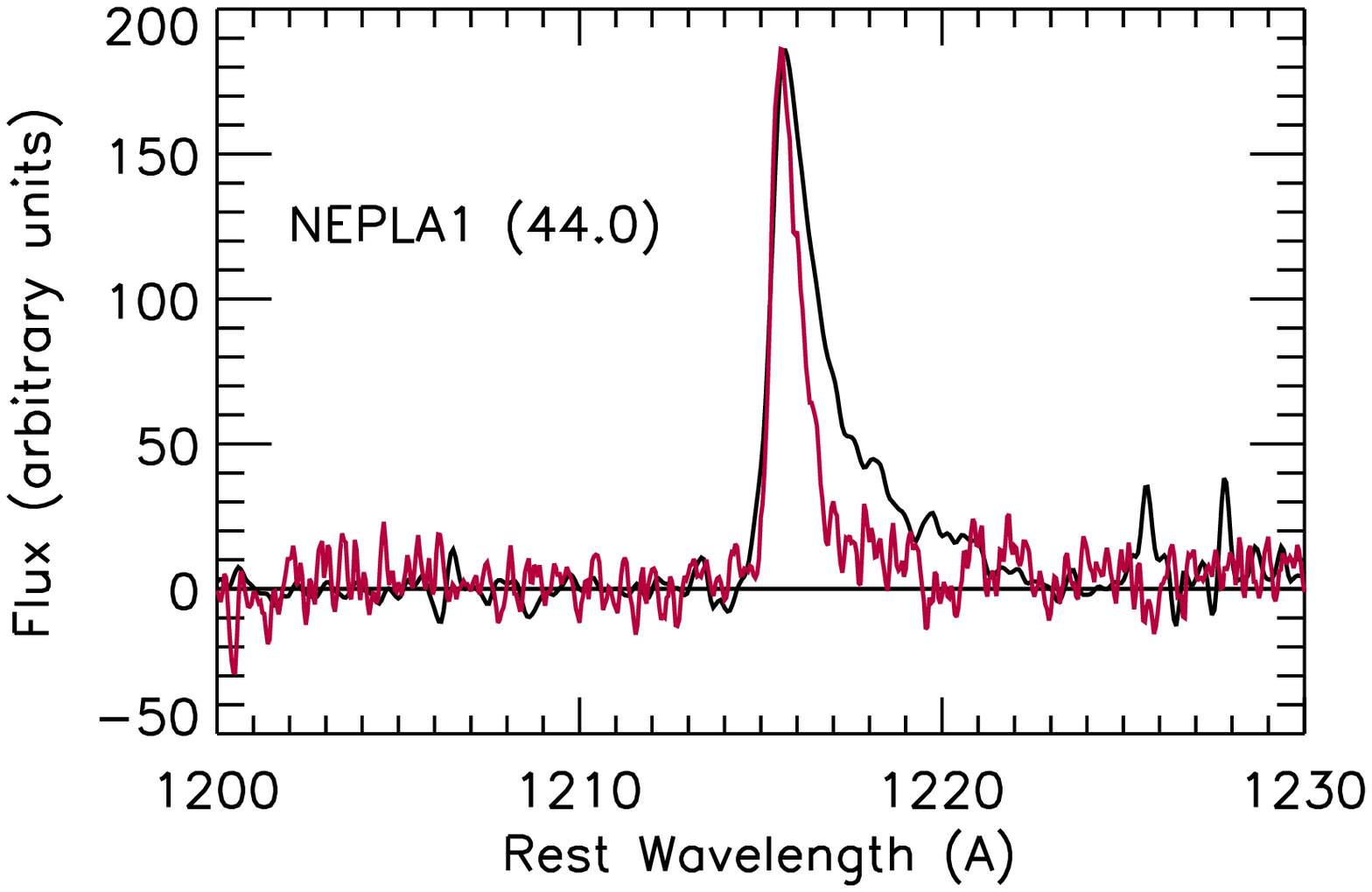}
\includegraphics[width=8.0cm,angle=0]{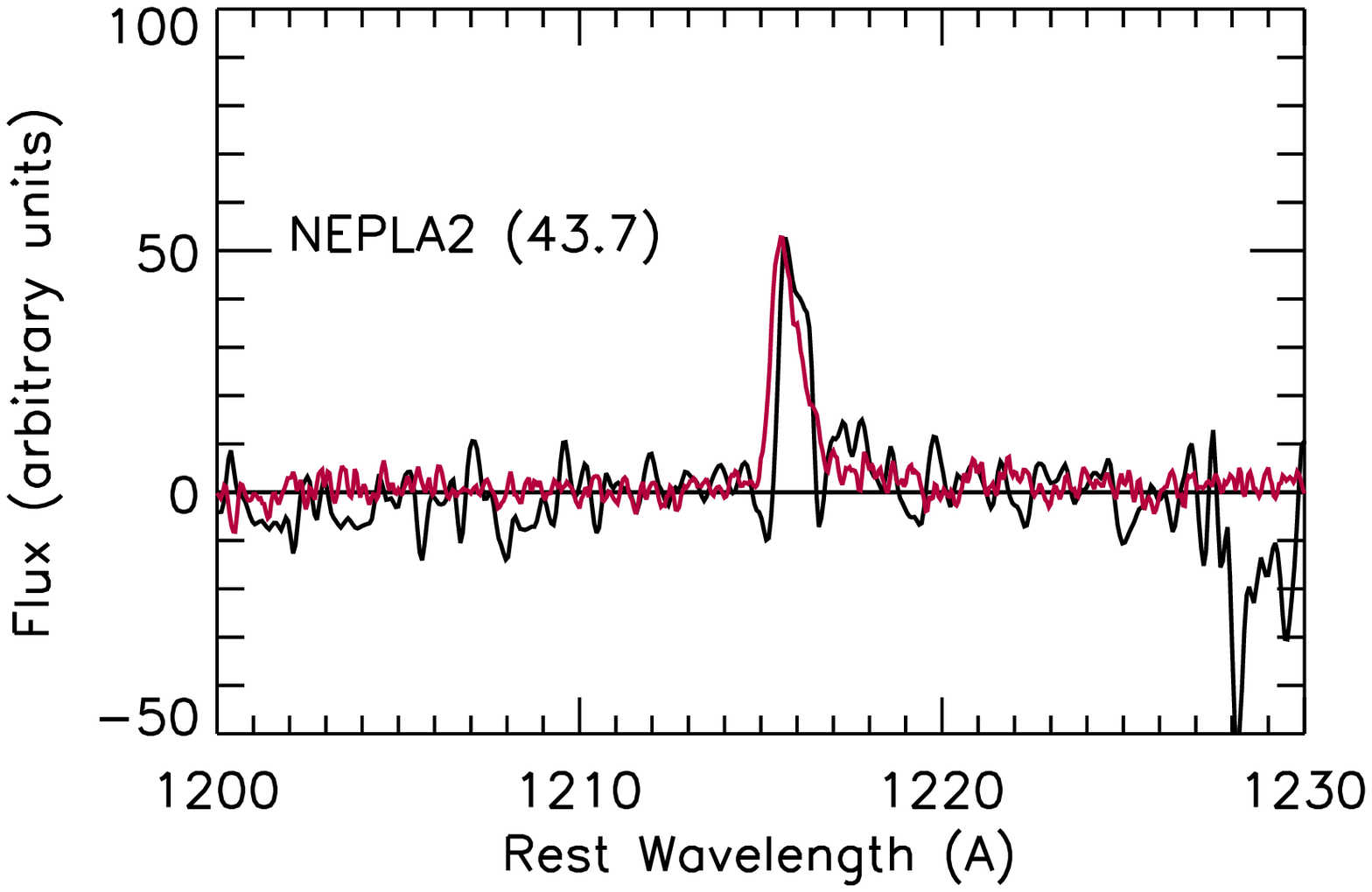}
\includegraphics[width=8.0cm,angle=0]{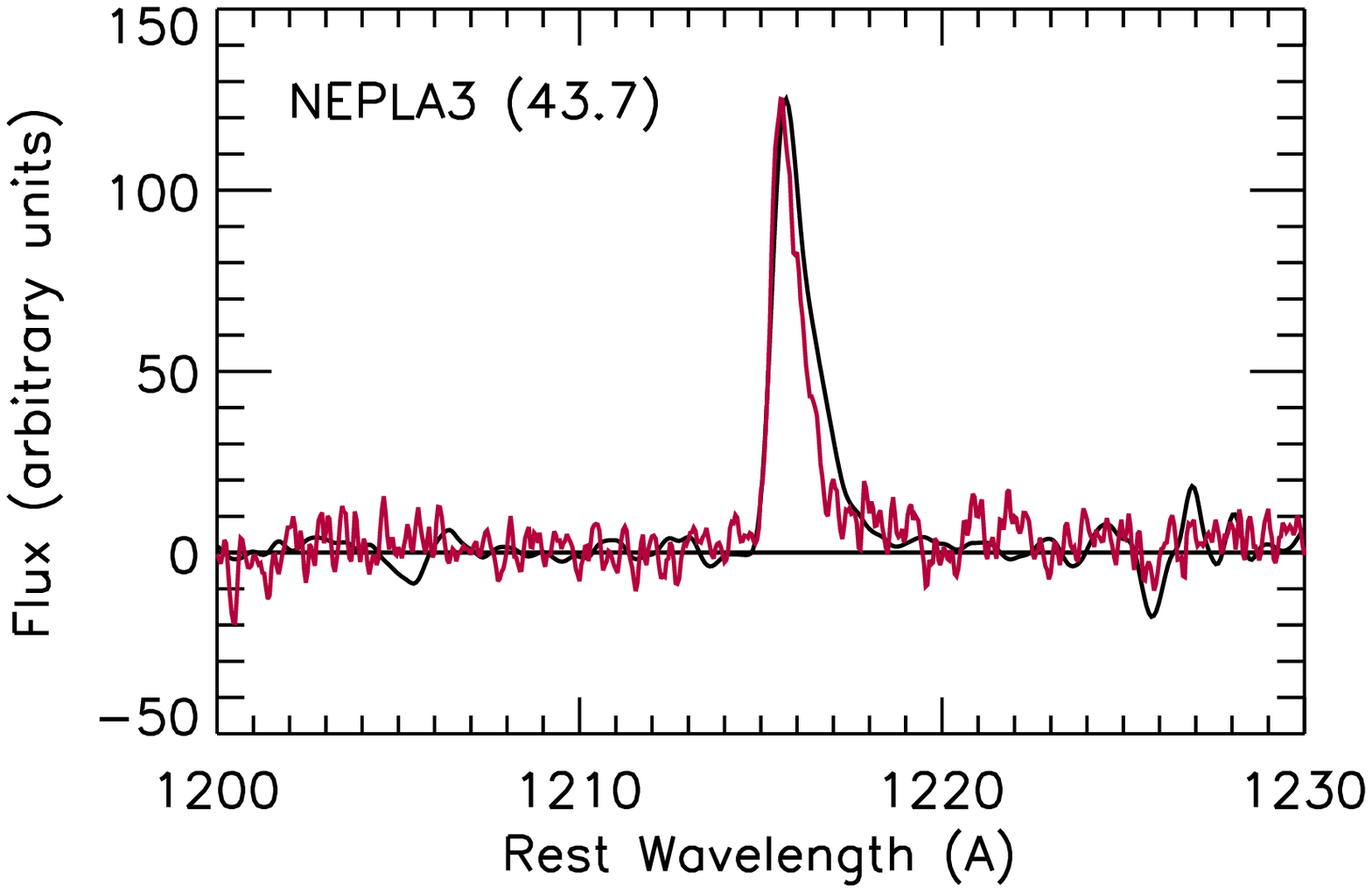}
\includegraphics[width=8.0cm,angle=0]{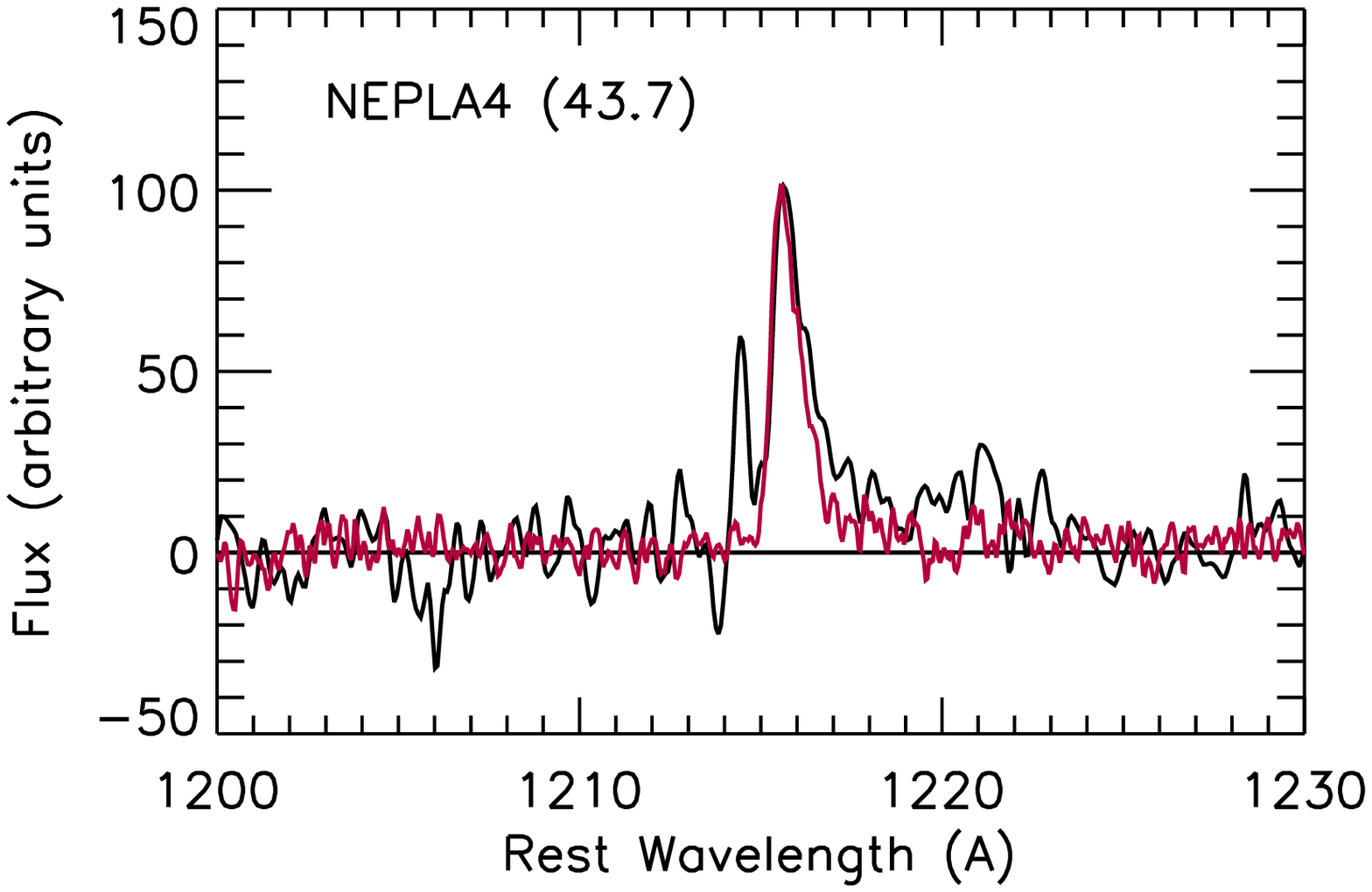}
\hspace{2.0cm}
\includegraphics[width=8.0cm,angle=0]{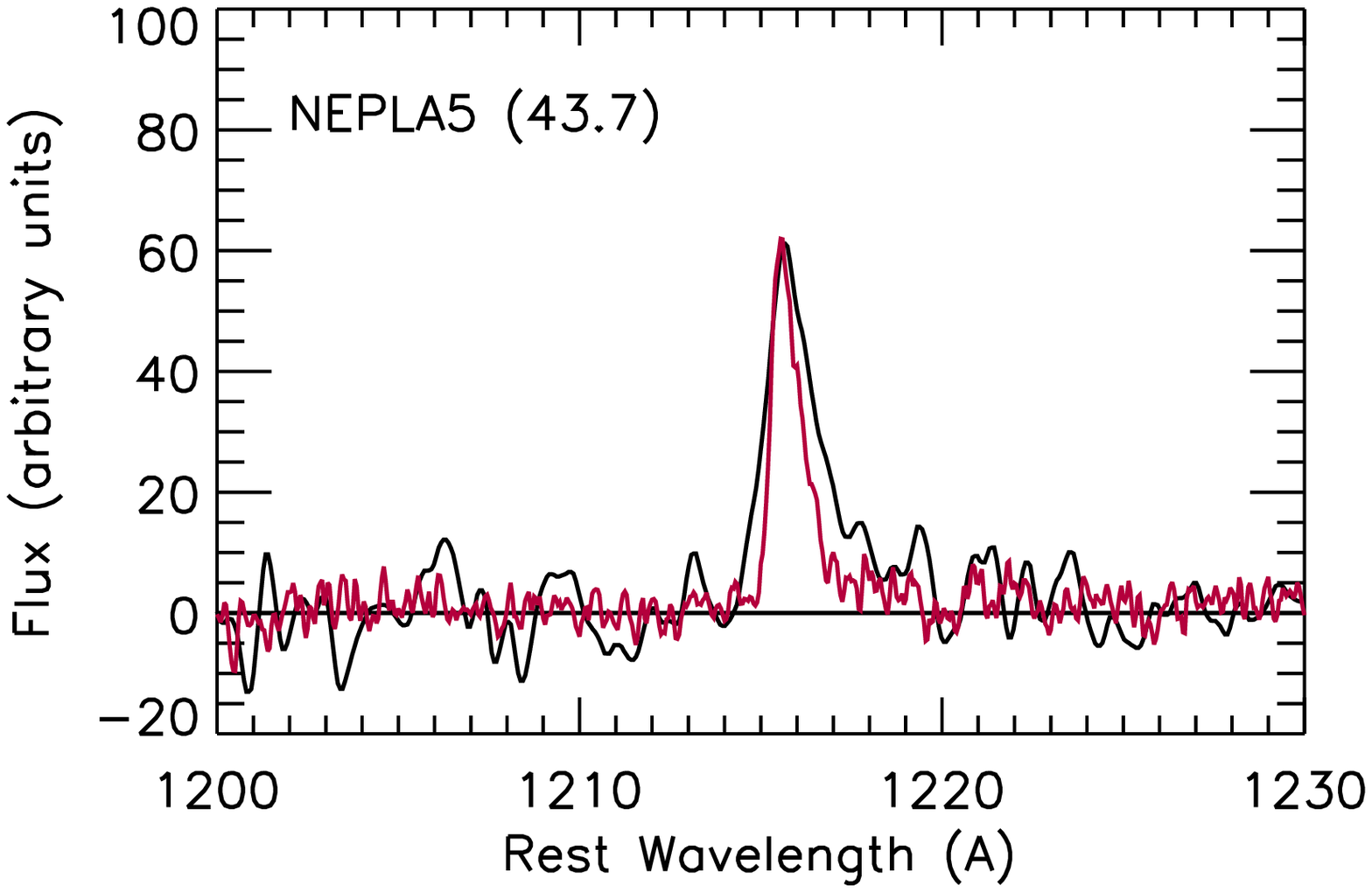}
\caption{Ly$\alpha$\ profiles of ultraluminous LAEs in the 
COSMOS and NEP fields (black), compared with the 
composite profile of 31 lower luminosity LAEs from
\citet{hu10} (red). The composite spectrum
is normalized to the peak flux in each spectrum. All spectra were 
obtained with DEIMOS on Keck~II using the 830~l/mm grating.  
See Table~\ref{tabobj} for details.
\label{spectra}
}
\end{figure*}

\begin{figure}[ht]
\centerline{\includegraphics[width=10cm,angle=0]{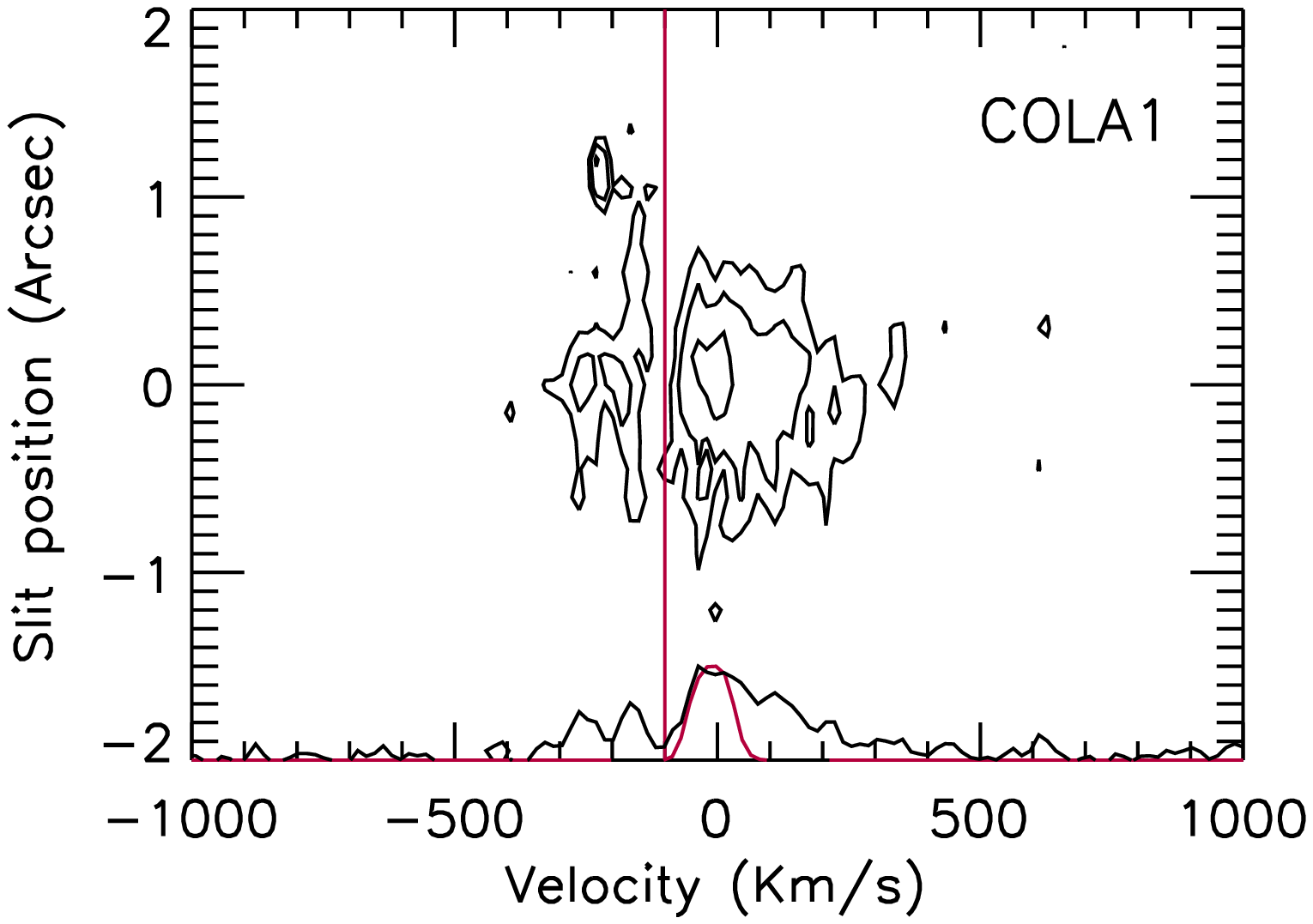}}
\centerline{\includegraphics[width=10cm,angle=0]{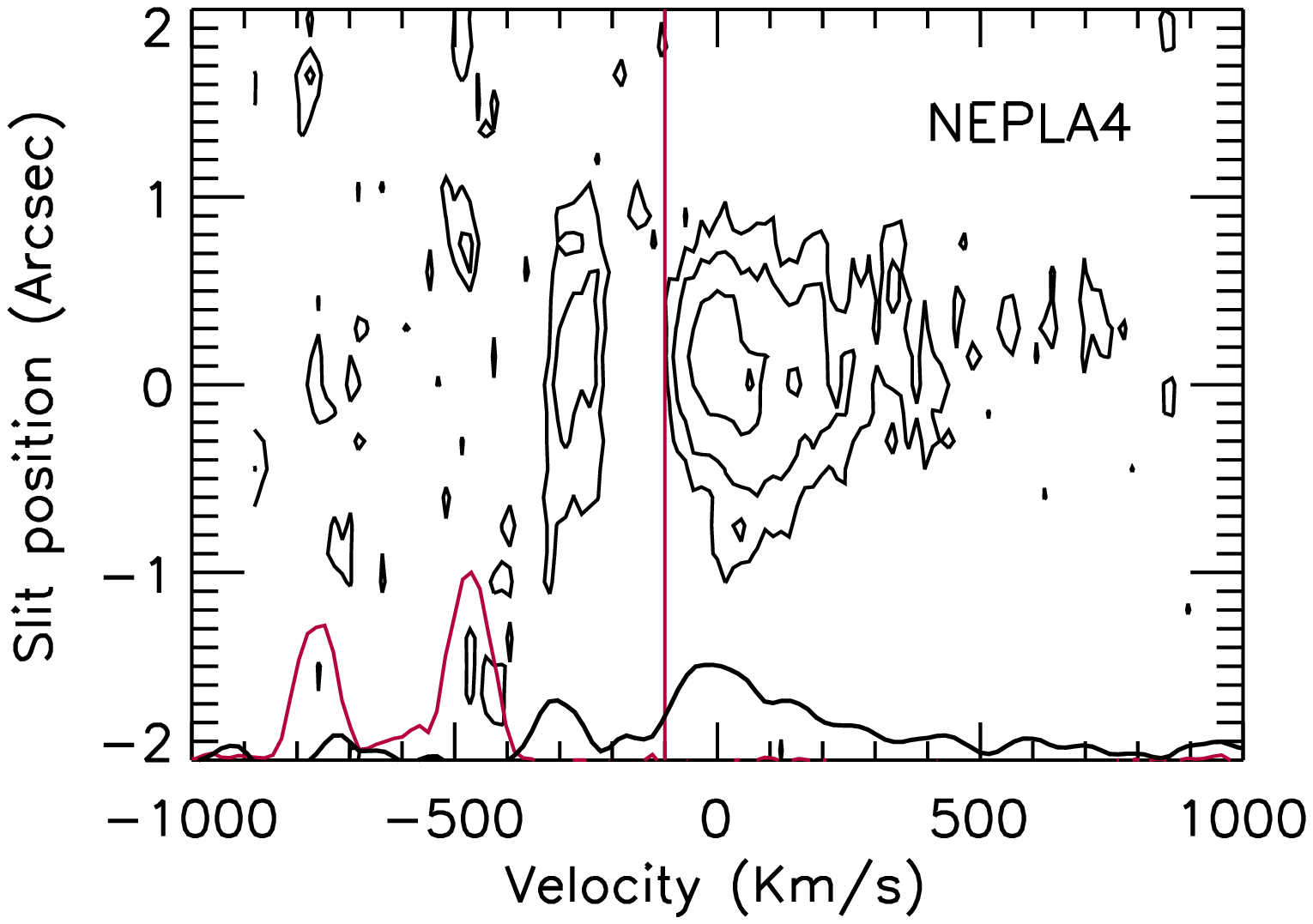}}
\caption{Ly$\alpha$ contours and profiles (black) for (top) COLA1 
and (bottom) NEPLA4. The maximum intensity in each case is shifted 
to zero velocity, and the red vertical line in each plot shows the edge
of the main structure. The red profiles at the bottom of each plot show 
the positions of the sky lines, and the black shows the spectrum.
\label{contours}
}
\end{figure}

\begin{figure}[ht]
\centerline{
\includegraphics[width=9.0cm,angle=0]{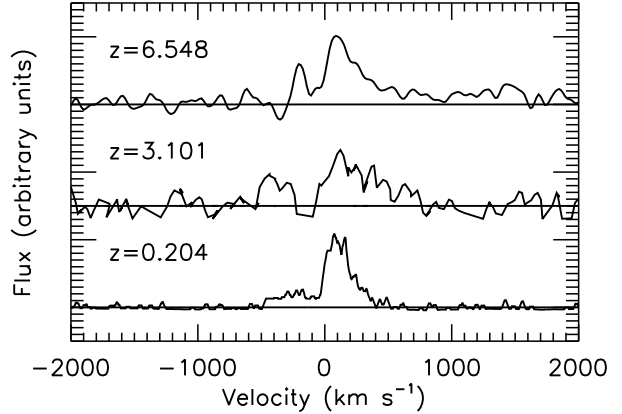}
}
\caption{Comparison of the spectral profile of NEPLA4 (top panel)
with a $z\sim3$ blue-wing LAE from \citet{yamada12} (middle panel) 
and a low-redshift blue-wing LAE from 
Scarlata et al.\ (2018, in preparation) (bottom panel). 
The low-redshift spectrum is shown 
in the systemic frame based on H$\alpha$ emission. For the two higher 
redshift sources, the velocity scale is chosen to match the peaks to
the low-redshift LAE peak.
\label{compare}
}
\end{figure}

Five of the seven observed candidates were identified as $z=6.6$
LAEs. We show their spectra in Figure~\ref{spectra},
along with those of CR7, MASOSA and COLA1\footnote{The observations
of NEPLA2 are of poorer quality, and we consider this
source more questionable than the other four NEP LAEs.}. The
measured redshifts corresponding to the peak of the
profiles are given in Table~\ref{tabobj}. In each case, we compare
the line profile (black) with the stacked spectra of lower luminosity
LAEs at $z=6.6$ ($\log L({\rm Ly}\alpha) < 43.3~{\rm erg\ s}^{-1}$)
from \citet{hu10} (red).
It is apparent from Figure~\ref{spectra} that most
of the lines in these ultraluminous galaxies are broader
than those in lower luminosity LAEs. In some cases, they have more extended
red wings.  

NEPLA4 has a complex profile with red and blue peaks and bears a 
strong resemblance to COLA1. In Figure~\ref{contours},
we compare the two-dimensional spectra of COLA1 and NEPLA4.
In NEPLA4, the blue component is at a more negative
velocity, and the red wing is more extended. In both cases,
the spatial profile of the blue wing along the slit matches that
of the red wing, suggesting we are seeing light from
a single source rather than from two overlapping galaxies
at slightly separated redshifts.

\begin{figure}
\centerline{
\includegraphics[width=9.0cm,angle=0]{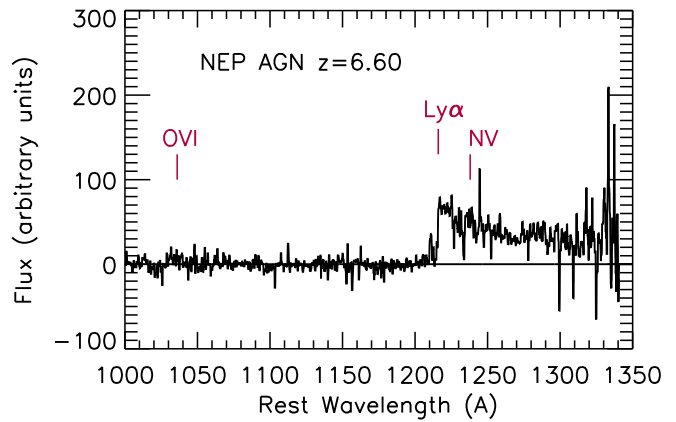}
}
\caption{Keck~II DEIMOS spectrum of the NEP AGN showing broad 
Ly$\alpha$ emission. See Table~\ref{tabobj} for details.
\label{nepagn}
}
\end{figure}

In Figure~\ref{compare}, we compare the spectral
profile of NEPLA4 with the profiles of two blue-wing LAEs at
lower redshift, one at $z=3.101$ from \citet{yamada12}
and one at $z=0.204$ from C.~Scarlata et al.\ (2018, in preparation).
NEPLA4 shows the same characteristic
double peak structure as the lower redshift
sources, with a strong dip between the peaks.
It also shows the same stronger and wider red peak.
In all cases, the red wing has an asymmetric profile extending
to higher velocities, which is characteristic of LAEs.
We will return to a quantitative analysis of the line profiles in 
Section~\ref{secphys}.

One of the remaining two sources is a very high equivalent width
[\ion{O}{3}]5007 emitter at $z=0.84$. The continuum in this source
is not detected in any of the blue bandpasses, making this
type of source hard to remove with a purely photometric
selection (see Figure~\ref{2d-images}). [\ion{O}{3}] emitters
are the primary contaminant and a significant source
of uncertainty in photometrically selected LAE
samples without follow-up spectroscopy 
\citep[e.g.,][and references therein]{konno17}.

The second source is a $z=6.6$ AGN (Figure~\ref{nepagn}). 
In this case, the Ly$\alpha$ forest break below the Ly$\alpha$
emission suppresses most of the light in the $z'$-band,
making this a narrowband-excess source. Sources
like this can be distinguished from LAEs if
the $y'$-band is sufficiently deep to measure the
longer wavelength continuum.

Luminosities were computed using the redshifts
and the NB921 and $z'$-band magnitudes. In Table~\ref{tabobj}, we give
the magnitudes measured in a $2{\arcsec}$ diameter aperture.
An average correction of 0.3~mag was applied to correct to total magnitudes. 
The spectrum was approximated with a single narrow line at the
observed Ly$\alpha$ wavelength and a flat $f_{\nu}$ 
continuum above this. We then folded this spectrum through
the transmission profiles for NB921 and $z'$ (see
Figure~\ref{921_filter}) and matched to the magnitudes
to determine the line flux and  continuum 
normalization. The luminosities were then computed 
from the line flux using the adopted cosmology, and are given
in Table~\ref{tabobj}. We have not corrected the Ly$\alpha$ fluxes
to allow for IGM scattering losses, since
it is unclear what correction to apply for these
sources. However, this could increase the luminosities
by as much as a factor of two.

\section{Contamination by Low-Redshift Emitters}
\label{lowemit}

\begin{figure}
\centerline{\includegraphics[width=10cm,angle=0]{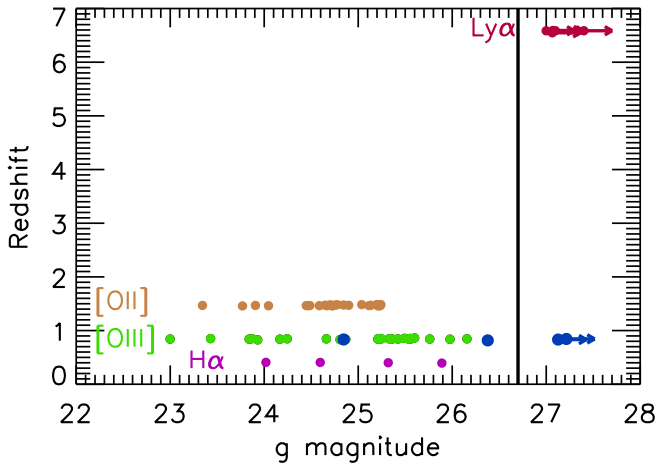}}
\caption{The $g'$ magnitude distribution for the lower redshift 
emitters chosen to have NB921$<23.5$ and $z - {\rm NB} > 0.7$, which 
selects H$\alpha$, [\ion{O}{2}]3727, and [\ion{O}{3}]5007 emitters, as compared
with 6 of the 7 primary LAE candidates, excluding only the high-redshift AGN.
The spectroscopically confirmed LAEs are shown in red, H$\alpha$ emitters in
purple, [\ion{O}{3}]5007 emitters in green, 
and [\ion{O}{2}]3727 emitters in gold. The vertical line shows
the $3\sigma$\ detection limit in $g'$. Only the [\ion{O}{3}]5007 emitters 
contain any sources that satisfy the LAE selection criterion $z' - {\rm NB} > 1.3$ 
(blue circles, including the one observed as part of the primary sample). 
\label{o2_gmg}
}
\end{figure}

We must be concerned about the possibility of confusion with lower redshift 
emission line galaxies for these high flux sources. In a purely photometric 
selection, the primary contaminants
are [\ion{O}{3}]5007 emitters, as we discuss below. However, with spectroscopic
follow-up, the [\ion{O}{3}]5007 emitters are easily identified, given the wide
doublet separation, and the main
worry becomes misidentifying lower redshift [\ion{O}{2}]3727 emitters as 
high-redshift LAEs. The reason for this is that the small doublet separation
in [\ion{O}{2}]3727 could produce 
spectra that look like complex Ly$\alpha$ profiles
if there are multiple velocity components in the emission
line galaxy.

Since we can  target only  one LAE candidate per DEIMOS mask,
we were able to explore more thoroughly the likelihood of contamination 
by strong emitters at low redshifts
by filling the remainder of each mask with sources having NB921$<23.5$ 
and $z'-$NB921 $>0.7$ but no other selection. 
In this way, we found 62 [\ion{O}{2}]3727 emitters, 
53 [\ion{O}{3}]5007, and 26 H$\alpha$ emitters. In Figure~\ref{o2_gmg}, 
we plot these, along with the 6 spectroscopically observed LAE candidates 
from our primary sample (Table~\ref{tabobj}, excluding only the high-redshift AGN)
versus $g^{\prime}$ magnitude. 
We show sources not detected in $g^{\prime}$ at the $2~\sigma$ upper limits
with right-pointing arrows. This includes all five spectroscopically
confirmed LAEs and two of the [\ion{O}{3}] emitters, including the one we 
observed as part of our primary sample.
The other $g^{\prime}$ faint [\ion{O}{3}] emitter is detected in the
combined $g^{\prime}$ + $r^{\prime}$ + $i^{\prime}$ data, and we did not 
include it in our primary sample. 

These extreme [\ion{O}{3}] emitters are the only sources, apart from the
LAEs, that are faint in $g^{\prime}$, and they
are the only group that includes sources that also satisfy the LAE selection 
criterion of $z'-$NB921 $>1.3$ (blue
circles in Figure~\ref{o2_gmg}).  However, they
are easily picked out with spectroscopy, because of the wide
doublet separation (as would also be the case for
\ion{C}{4} or \ion{Mg}{2} emitters). 

The [\ion{O}{2}]3727 emitters are all easily detected in $g^{\prime}$. 
In contrast, the LAE with a complex profile, NEPLA4, is not detected
in $g^{\prime}$ (Figure~\ref{contours}) with a $2\sigma$ limit of 27.1.
It is also redder than $z^{\prime}-$NB921 $>1.3$, which none of the
[\ion{O}{2}] emitters reach. It would therefore be highly anomalous 
photometrically if the line we identified as Ly$\alpha$ were instead
[\ion{O}{2}]3727.

\begin{figure*}[t]
\centerline{\includegraphics[width=15cm,angle=0]{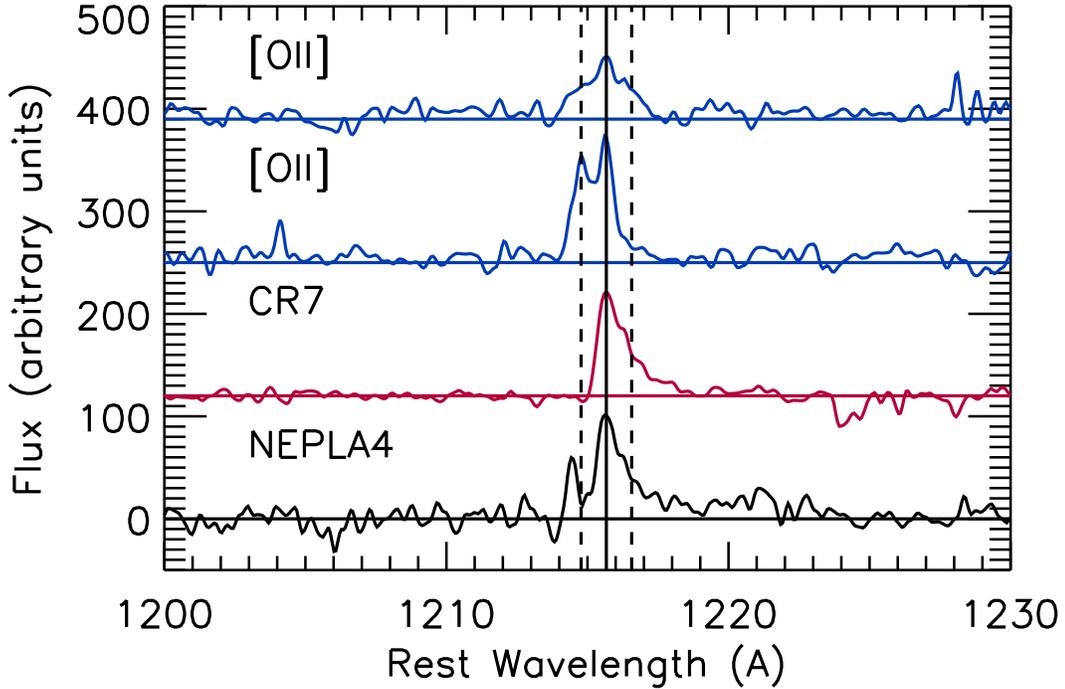}}
\caption{Spectral profiles from DEIMOS on Keck~II of (from bottom) 
NEPLA4, CR7, a ``standard" [\ion{O}{2}] emitter, and an [\ion{O}{2}] emitter
with two velocity components, giving it a more complex profile. All of the
profiles are shown at the rest wavelength they would have if the
line is Ly$\alpha$. The peak of the profile has been placed at zero velocity 
(solid vertical line). 
For the [\ion{O}{2}] emitters (top two panels), the dashed vertical lines 
show the position of the second member of the doublet, depending on
which member of the doublet is assumed to correspond to the peak.
\label{compare_o2}
}
\end{figure*}

However, we can make an even more compelling argument against
the line being [\ion{O}{2}]3727 using
the shape of the emission line profile. In Figure~\ref{compare_o2},
we show the line profiles of NEPLA4 and CR7, and, for comparison,
of two very strong [\ion{O}{2}]3727 emitters in the field.  
One of these [\ion{O}{2}] emitter profiles is a simple doublet, while the 
other has a more complex profile with two velocity components. The latter 
is the type of source most likely to be confused with a high-redshift 
blue-wing LAE. Since we do not know the systemic velocities of the 
galaxies, we measure the redshifts at the peaks of the profiles and assume
that they are at zero velocity (solid vertical line). For the [\ion{O}{2}] emitters 
(top two panels), we draw two dashed vertical lines to show 
where the second member of the doublet would lie, depending on which 
member of the doublet is assumed to correspond to the peak. 

We can see 
right away that the velocity separation between the blue wing and the
peak in NEPLA4 is larger than the [\ion{O}{2}]3727 doublet separation
(i.e., the difference between the solid line and one of the dashed lines). 
Furthermore, NEPLA4 has the strong red tail characteristic of LAEs 
(see, e.g., the CR7 spectrum in Figure~\ref{compare_o2}), and this is 
not seen in the [\ion{O}{2}]3727 emitters.
We conclude that NEPLA4 is not a lower redshift [\ion{O}{2}] emitter.

\section{Line structure}
\label{secline}
From Figure~\ref{spectra}, we can see that the line profiles
of the ultraluminous LAEs are wider than those of the
lower luminosity LAEs, and that, in two cases, they have complex
profiles (COLA1 and NEPLA4). (MASOSA also has structure in its
red tail). If these are interpreted as
blue-wing LAEs, then roughly one-third of the ultraluminous LAEs
fall into this category. However, with the present small numbers,
the fraction is quite uncertain, with a 68\% confidence range from
0.12 to 0.77.

\begin{figure}[H]
\centerline{\includegraphics[width=7cm,angle=0]{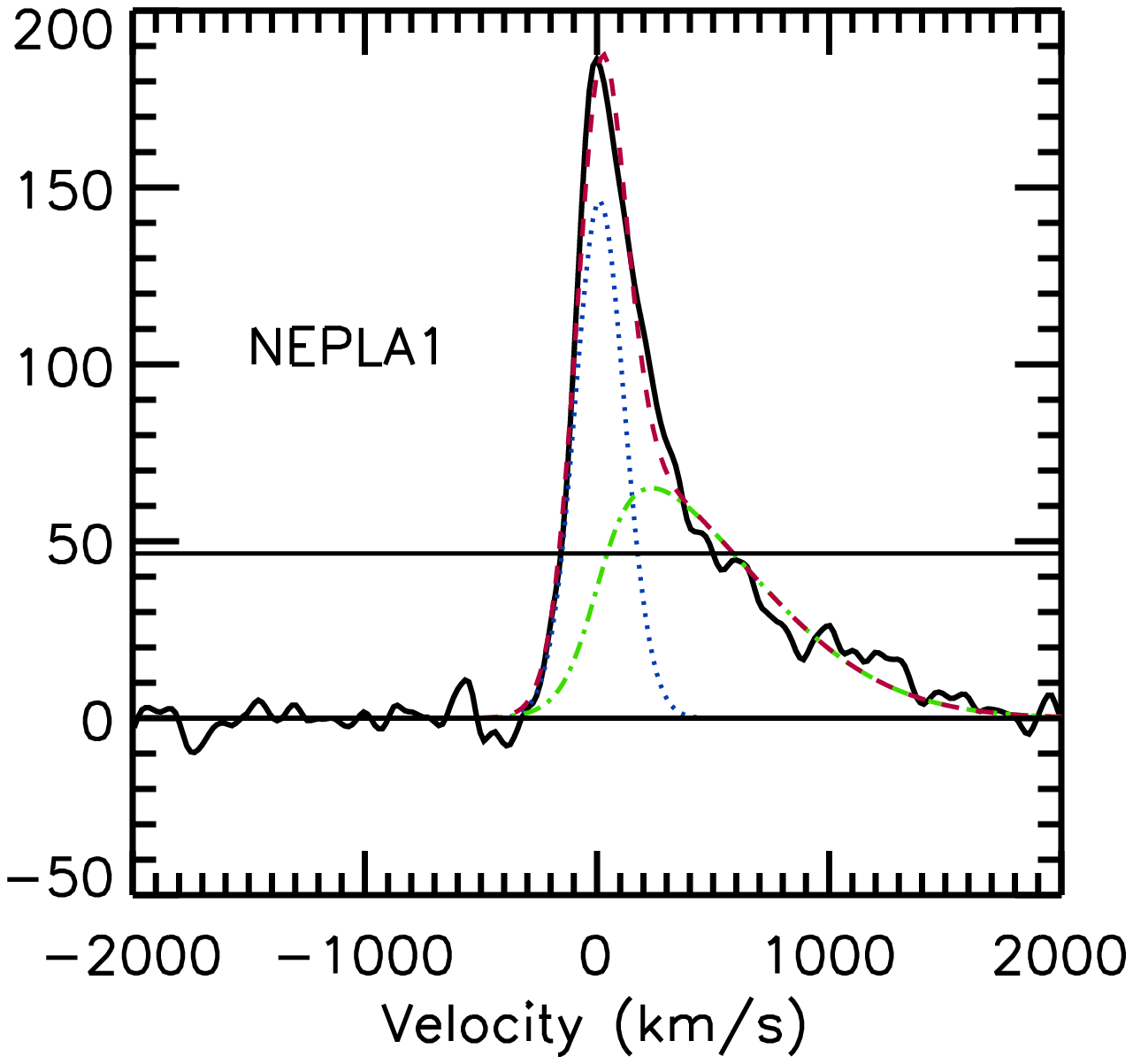}}
\centerline{\includegraphics[width=7cm,angle=0]{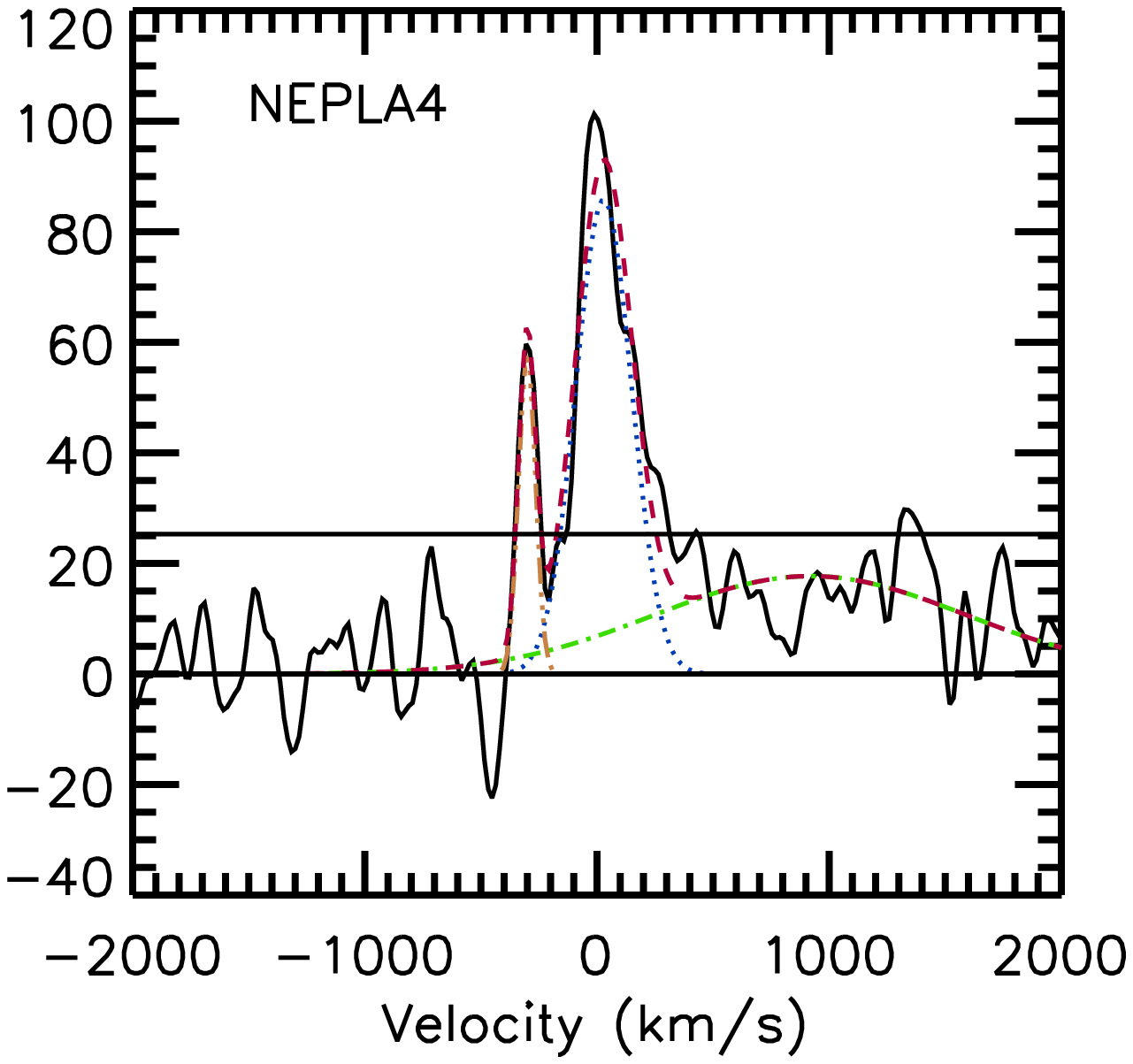}}
\caption{Ly$\alpha$ profiles (black) of NEPLA1 (top) and NEPLA4 (bottom).  
Gaussian fits to the main profiles (blue dotted), to the 
red tails (green dot-dashed), and, for NEPLA4, to the blue
wing (gold dashed) are also shown, as are the combined fits (red dashed). 
The black lines show the FWQM levels. The Gaussian fitting 
parameters are given in Table~2.
\label{gauss-fits}
}
\end{figure}

In \citet{hu10}, we fitted the spectral profiles with a single
demi-Gaussian (a Gaussian cut away on the blue side) convolved
through the instrument profile. However, the present spectral profiles 
are more complex, so we have instead fitted them with either two
(for the non-blue-wing LAEs) or three (for the blue-wing LAEs) Gaussians. 
We provide these fits in Table~2
where we give the amplitude, 
central velocity, and $\sigma$ for each component. 
Again, we choose the zero velocity
to correspond to the peak of the profile. We show two examples
of the fits in Figure~\ref{gauss-fits}.

For the non-blue-wing LAEs, we also fitted the spectral profiles with 
two truncated Gaussians with cutoffs below the zero velocity. 
We provide the parameters of these fits in Table~2 
in the lines labeled ``cutoff". All of these fits make assumptions
about the properties of the lines, so we have also given two
simple measures of the line widths---the full-width at half-maximum
(FWHM) and the full-width at quarter-maximum (FWQM)---in columns
(2) and (3). For the FWQMs, we restricted to velocities within 
$\pm 1000~{\rm km\ s}^{-1}$ of the peak.

In order to compare with the lower luminosity LAE sample, we measured 
the FWQMs for the $z=6.5-6.6$ LAEs in \citet{hu10} with high enough S/N 
for this quantity to be well defined. We also measured the FWQM for the 
stacked spectrum of all quality~1 sources in \citet{hu10};  
we provide these results and fits as the last entries in Table~2.

\begin{figure}[ht]
\centerline{\includegraphics[width=9.3cm,angle=0]{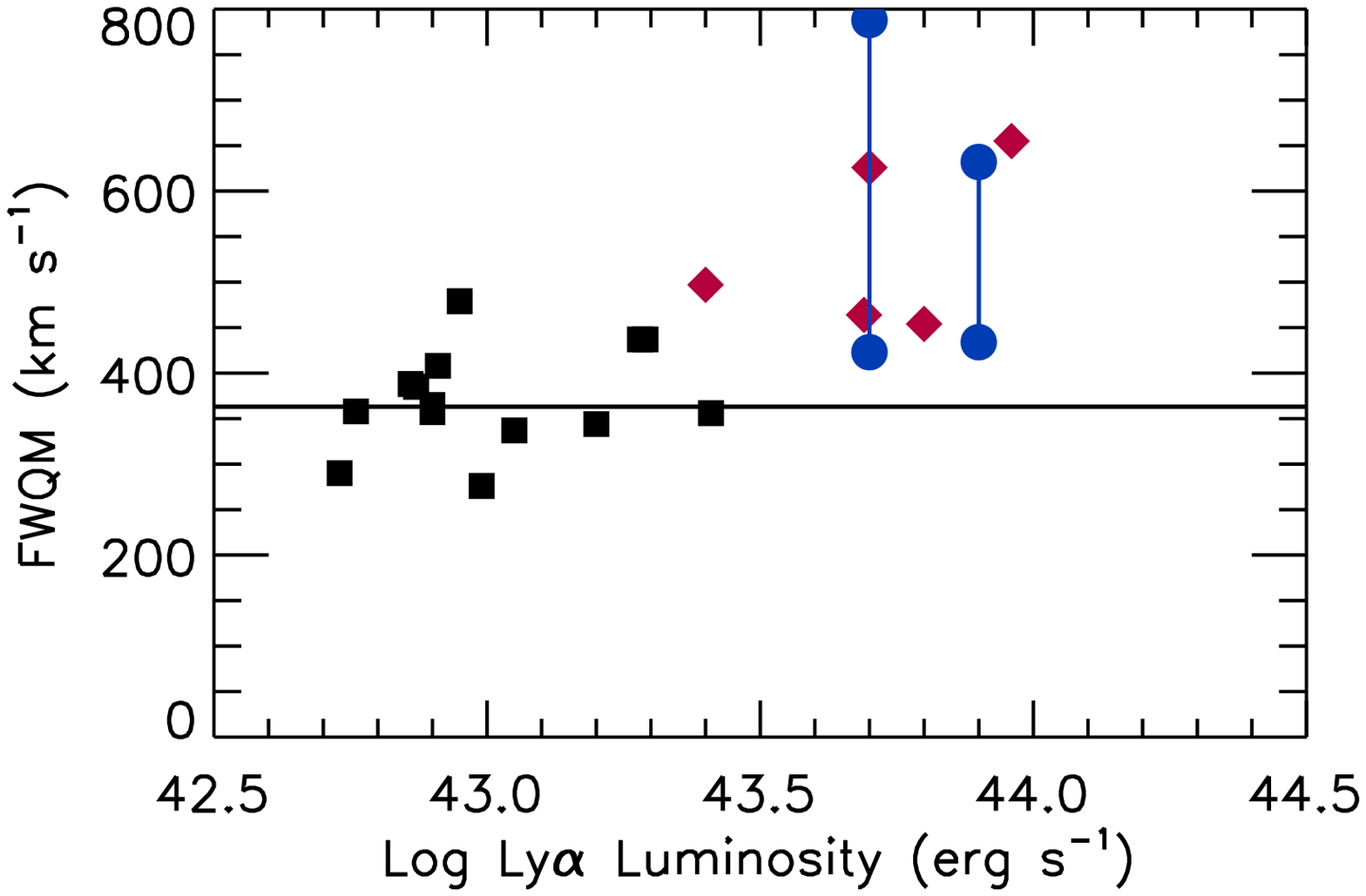}}
\centerline{\includegraphics[width=9.3cm,angle=0]{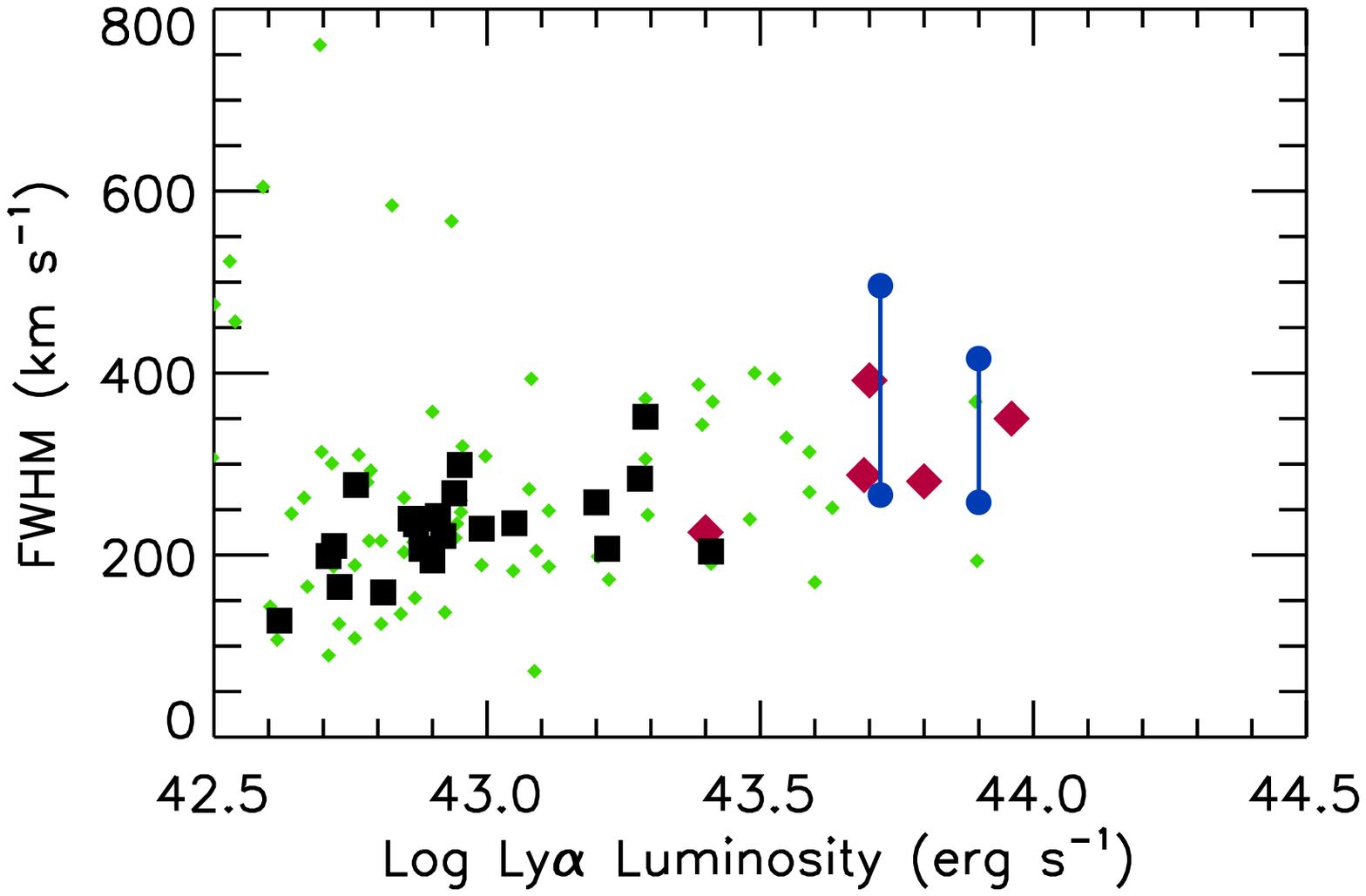}}
\caption{(Top) Comparison of the FWQM of the
present sample with the lower luminosity sample of
\citet{hu10}. Black squares show individual
measurements of the lower luminosity sample,
and the black horizontal line shows the measurement
in the stacked spectrum. The measurements in the
present sample (Table~2) are shown
in red if there is no blue wing, and as blue circles
if there is a blue wing (upper point). For the blue wing cases,
we also show the FWQM measured excluding
the blue wing (lower point). (Bottom) FWHM of the present
sample with the same notations as above. Here we compare
both with the \citet{hu10} sample (black squares) and
with the compilation of \citet{matthee17b} 
(green diamonds).
\label{fwq_lum}
}
\end{figure}

In Figure~\ref{fwq_lum}, we compare the FWQMs for the lower luminosity 
LAE sample (black squares for the individual points, and black line for the 
stacked spectrum) with those from the ultraluminous LAE sample 
(red diamonds for the non-blue-wing LAEs, and blue circles for the
blue-wing LAEs). We find those for the lower luminosity LAE sample to be 
systematically lower. For the blue-wing LAEs, we show both the FWQM 
(upper point) and the result once we exclude the blue wing (lower point).

The transition between the relatively uniform lower luminosity LAE
sample and the wider range of the ultra-luminous
LAE sample corresponds very roughly to 
$\log {L({\rm Ly}}\alpha) = 43.5~{\rm erg\ s}^{-1}$,
though with the present relatively small sample
the transition point is quite uncertain, at least
at the 0.1-0.2 dex level (see Figure~\ref{fwq_lum}).
However, the difference between the high- and low-luminosity 
 samples is highly significant.
A Mann-Whitney test give a two tailed probability
of 0.0027 for the $\log {L({\rm Ly}}\alpha) > 43.5~{\rm erg\ s}^{-1}$
LAEs being drawn from the same sample as the $\log {L({\rm Ly}}\alpha) < 43.5~{\rm erg\ s}^{-1}$ LAEs.
If we divide instead at 43.4, then the two tailed probability is 0.0034.

This difference is le thenss clearly seen in the FWHMs,
which we show in the lower panel of Figure~\ref{fwq_lum}.
In this panel, in addition to comparing with the Hu et al.
(2010) sample (black squares) we also 
compare with the compilation of \cite{matthee17b}
(excluding CR7 and MASOSA, which otherwise would appear twice). 
The Matthee compilation shows more scatter, possibly
partly reflecting more variation in the quality of the
spectra, but both this and the Hu sample show a somewhat smoother
rise in the FWHM widths as we move to higher luminosities
than is seen in the FWQMs.
It appears that the changes in the profiles are mostly in
the red tails (see Figure~\ref{spectra}), and the FWQMs provide a better 
characterization of the red tails than the FWHMs.

\begin{deluxetable*}{lllcccll}
\tablecaption{Properties of Primary Sample with Spectroscopic Follow-up\label{tabobj}}
\tablewidth{400pt}
\tablenum{1}
\tablehead{
\colhead{R.A. (2000)} &
\colhead{Decl. (2000)} &
\colhead{NB921} &
\colhead{$z'$-band} &
\colhead{Flux\tablenotemark{a}} & 
\colhead{$\log L(\rm Ly\alpha$)\tablenotemark{a}} &
\colhead{Redshift} & 
\colhead{Name}\\
\colhead{(deg)} &
\colhead{(deg)} &
\colhead{(AB)} &
\colhead{(AB)} &
\colhead{(erg cm$^{-2}$ s$^{-1}$)} &
\colhead{(erg s$^{-1}$)} &
\colhead{} &
\colhead{} \\
\colhead{(1)} & \colhead{(2)} & \colhead{(3)} & \colhead{(4)} & \colhead{(5)} & \colhead{(6)} &
\colhead{(7)} & \colhead{(8)}
}
\startdata
   273.73837  & 65.285995  &  22.75  &  25.16  &  1.9$\times 10^{-16}$  & 43.96  & 6.5942   & NEPLA1 \cr
   263.17966  & 65.520416  &  22.91  &  24.39  &  \dots  & \dots  & 6.6000   & NEPAGN \cr
   277.74066  & 68.367943  &  23.17  &  24.57  &  \dots  & \dots  & 0.8439  &  OIII \cr
   263.61490  & 67.593971  &  23.20  &  25.31  &  1.2$\times 10^{-16}$ & 43.76  & 6.5835?  & NEPLA2 \cr
   265.22437  & 65.510361  &  23.42  &  25.23  &  1.0$\times 10^{-16}$  & 43.69  & 6.5915   & NEPLA3 \cr
   268.29211  & 65.109581  &  23.30  &  25.42  &  1.1$\times 10^{-16}$  & 43.72  & 6.5480   &  NEPLA4 \cr
   269.68964  & 65.944748  &  23.47  &  24.95  &  1.0$\times 10^{-16}$  & 43.70  &  6.5367 & NEPLA5 
\enddata
\tablenotetext{a}{Fluxes and luminosities are only given for confirmed $z=6.6$ LAEs.}
\end{deluxetable*}

\begin{deluxetable*}{cccccccccccc}
\tablecaption{Parameters of Gaussian Fits\label{tabfits}}
\tablewidth{0pt}
\tablenum{2}
\tablehead{
\colhead{} & \colhead{} & \colhead{} & \multicolumn{3}{c}{Component~1} & \multicolumn{3}{c}{Component~2} &\multicolumn{3}{c}{Component~3} \\
\colhead{} & \colhead{} & \colhead{} & \multicolumn{3}{c}{(main)} & \multicolumn{3}{c}{(red tail)} &\multicolumn{3}{c}{(blue wing)} \\
\colhead{} & \colhead{} & \colhead{} & \colhead{} & \colhead{} & \colhead{} & \colhead{} & \colhead{} & \colhead{} & \colhead{} & \colhead{} & \colhead{} \\
\cline{4 - 6}
\cline{7 - 9}
\cline{10 - 12}
\colhead{} & \colhead{} & \colhead{} & \colhead{} & \colhead{} & \colhead{} & \colhead{} & \colhead{} & \colhead{} & \colhead{} & \colhead{} & \colhead{} \\
\colhead{} & \colhead{FWHM} & \colhead{FWQM} & 
\colhead{amp} & \colhead{$v_c$} & \colhead{$\sigma$} &
\colhead{amp} & \colhead{$v_c$} & \colhead{$\sigma$} & 
\colhead{amp} & \colhead{$v_c$} & \colhead{$\sigma$} \\
\colhead{} & \colhead{(${\rm km\,s}^{-1}$)} & \colhead{(${\rm km\,s}^{-1}$)} & 
\colhead{} & \colhead{(${\rm km\,s}^{-1}$)} & \colhead{(${\rm km\,s}^{-1}$)} &
\colhead{} & \colhead{(${\rm km\,s}^{-1}$)} & \colhead{(${\rm km\,s}^{-1}$)} &
\colhead{} & \colhead{(${\rm km\,s}^{-1}$)} & \colhead{(${\rm km\,s}^{-1}$)} \\
\colhead{(1)} & \colhead{(2)} & \colhead{(3)} & \colhead{(4)} & \colhead{(5)} & \colhead{(6)} &
\colhead{(7)} & \colhead{(8)} & \colhead{(9)} & \colhead{(10)} & \colhead{(11)} & \colhead{(12)}
}
\startdata
CR7     &  281  &  454   &     86   &  -3   &  68   &  53   & 175   & 154  & \dots  &  \dots  & \dots \cr
cutoff  &  281  &  454   &    181   & -39   &  67   &  76   &  32   & 228  &  \dots  &  \dots  & \dots \cr
&&&&&&&&&&& \cr
MASOSA  &  225  &  497   &     82   & -19   &  80   &  32   & 219   & 150  & 	 \dots  &  \dots  & \dots \cr
cutoff  &  225  &  497   &    241   & -77   &  92   &  34   & 213   & 150  &  \dots  &  \dots  & \dots \cr
&&&&&&&&&&& \cr
NEPLA3  &  288  &  464   &     95   &  -5   &  79   &  56   & 169   & 154  & 	 \dots  &  \dots  & \dots \cr
cutoff  &  288  &  464   &    219   & -60   &  83   &  86   &  28   & 216  &  \dots  &  \dots  & \dots \cr
&&&&&&&&&&& \cr
NEPLA1  &  350  &  655   &    153   &  32   & 122   &  52   & 443   & 36   &  \dots  &  \dots  & \dots \cr
cutoff  &  350  &  655   &    378   &-100   & 138   &  73   &   0   & 613  & \dots  &  \dots  & \dots \cr
&&&&&&&&&&& \cr
NEPLA5  &  392  &  626   &     50   &  34   & 152   &  10   & 520   & 516  &  \dots  &  \dots  & \dots \cr
cutoff  &  392  &  626   &    131   &-100   & 145   &  19   &   0   & 658  & \dots  &  \dots  & \dots \cr
&&&&&&&&&&& \cr
COLA1   &  416  &  632   &     70   &   8   &  69   &  55   &  67   & 236  &   40  & -242  &   594 \cr
&&&&&&&&&&& \cr
NEPLA4  &  496  &  788   &     85   &  28   & 120   &  17   & 919   & 667  &   57  & -300  &   36\cr
&&&&&&&&&&& \cr
Lower Lum. &&&&&&&&&&& \cr
Stacked    &   221  &   366    &     15  &   -15   &   66    &   9    & 136  & 150   & \dots & \dots & \dots \cr
cutoff   &    221  &   366    &     38   &  -52  &    68    &  14   &    0    &   209 & \dots & \dots & \dots	  	  
\enddata
\end{deluxetable*}

\section{Discussion}
\label{secphys}
\subsection{Complex line profiles}

In contrast to the lower luminosity LAEs at $z=6.6$, which show
very little variation in either the line profiles or the line
widths \citep{hu10}, we have found that the LAEs at 
$\log L({\rm Ly}\alpha) > 43.5~{\rm erg\ s}^{-1}$
show both wider and more complex profiles.
At lower redshifts ($z= 2-3$), \citet{konno15} and \citet{matthee17b} 
suggest that most $\log L({\rm Ly}\alpha) > 43.4~{\rm erg\ s}^{-1}$ LAEs
are associated with AGNs. However, at $z=6.6$, Matthee et al.\ argue that
the relative narrowness of the Ly$\alpha$ lines
(see \citealt{alexandroff13,matsuoka16}) and the absence of 
\ion{C}{4} 1550~\AA\ suggest that the sources may primarily be powered 
by star formation rather than by AGN activity. We note that COLA1 also 
shows no sign of CIV in a deep MOSFIRE observation of the $J$-band 
(A.~Barger et al.\ 2018, in preparation). Nevertheless, we must keep in mind
the possibility that the ultraluminous LAEs may have contributions 
to their emission from AGNs.

While the wider profiles in the ultraluminous LAEs might be understood as 
being caused by higher velocity outflows in these more luminous
galaxies, the presence of the more complex profiles suggests
that a more likely explanation is that these LAEs lie in highly ionized regions. 
This allows the full Ly$\alpha$ emission line profile of
the galaxy itself to be visible. The absence of IGM scattering near
the LAE would both increase the observed width of the line and
allow the blue wing to be visible.

Another line of evidence that ultraluminous LAEs may lie in different 
environments  comes from
\citet{matthee15} and \cite{bagley16}, who find little evolution in the
ultraluminous LAE LF between $z=5.7$ and $z=6.6$ and between $z=6.6$
and $z=6.9$, respectively. In contrast, the lower
luminosity ($\log L({\rm Ly}\alpha)<43.5$~erg~s$^{-1}$) LAE LF evolves
significantly (by a factor of $\sim 2$ in normalization)
over the redshift range $z=5.7$ and $z=6.6$
\citep{hu10,ouchi10}. Matthee et al.\ and Bagley et al.\
argue that if the evolution of the
lower luminosity LAE LF is at least partly driven by the increasing
neutrality of the IGM with increasing redshift, then a relative lack of evolution
in the ultraluminous LAE LF could be due to these sources lying
in large \ion{H}{2} regions that protect them from changes in the IGM
neutrality. However, we caution that the normalization of the 
ultraluminous LAE LF at $z=6.6$ used by Matthee et al.\ has been 
questioned by \cite{konno17}.

Unfortunately, understanding Ly$\alpha$ profiles is not
easy at any redshift \citep[e.g.,][]{verhamme12}.  
In particular, modeling Ly$\alpha$ profiles requires making assumptions 
about the underlying galaxy and the amount and phase structure of outflows
and inflows (although the kinematics of the high-redshift LAEs may be explored
directly with ALMA observations of the [\ion{C}{2}] fine structure
line, as \citealt{matthee17a} 
have shown for CR7). Thus, rather than trying to compare with models,
we instead compare with lower redshift LAEs. This also involves
assumptions, since the galaxy properties and the relative
importance of outflows and inflows may change with redshift, 
but it is relatively straightforward.

Based on the present NEP observations and the previous data from
the COSMOS field \citep{hu16}, we estimate
that the fraction of blue-wing LAEs in the $z=6.6$ ultraluminous
LAEs is $33_{12}^{77}$\%, where the subscript and
superscript give the 68\% confidence range. With the
present relatively small number of sources, the uncertainty
is large, but this is broadly consistent with the
results at lower redshifts, where about 30\% of
LAEs at $z=2-3$ show multi-component profiles \citep{kulas12}. 
(\citealt{yamada12} give an even higher fraction of around 50\%
at $z=3$, but this may be  biased upwards owing to the noise
in the spectra. The Yamada region is also significantly
overdense and this might also affect the properties
of the LAEs.) Scarlata et al.\ (2018, in preparation)
see a similar fraction of about 30\% blue-wing LAEs in a 
$z=0.3$ LAE sample observed with the COS spectrograph on {\em HST\/}.
We do not expect these $z=0.3$ LAEs to be strongly
affected by IGM opacity, so this percentage 
could be applicable to high-redshift galaxies enclosed in
giant \ion{H}{2} regions, if the underlying galaxy properties are similar.

Since we do not have measurements of other optical
or UV lines, we cannot establish an absolute velocity
scale for the Ly$\alpha$ lines. However, the variety
of profiles appears very similar to what is seen in
LAEs at lower redshift.  Even for the pure red-tail
sources, such as NEPLA1 and CR7, there are many similar
low-redshift sources with red tails and sharp cutoffs on
the blue side with which to compare; this type of profile can
therefore arise without the presence of IGM scattering.

For the complex profile LAEs (i.e., NEPLA4 and COLA1),
we measure the peak-to-peak velocity separations
between the main and blue wing components to be
300 and 270~${\rm km\ s}^{-1}$, respectively.
These are smaller than the values seen in the $z=2-3$ sample,
where \citet{kulas12} find a median peak-to-peak
separation of 757~${\rm km\ s}^{-1}$ in the multi-peak sources, but similar
to the values seen in the $z=0.3$ sample of Scarlata et al.\ (2018),
where the median peak-to-peak separation is 339~${\rm km\ s}^{-1}$. 
This might suggest that the $z=0.3$ LAEs are better analogs to
the $z=6.6$ LAEs in terms of their inflow and outflow properties.

Both NEPLA4 and COLA1 have Ly$\alpha$ profiles with strong blue sides:  
the ratios of the peak of the blue wing component to the peak of the
red tail component are 0.6 and 0.5, respectively. These are stronger
than those generally seen in low-redshift samples. The
mean ratio in the Scarlata et al.\ (2018) $z=0.3$
sample is 0.35, though the maximum value is 0.66. The strong blue sides
at $z=6.6$ suggest that we must be seeing significant infall.
This, in turn, suggests that these may be young galaxies
in the early stages of formation and consistent with 
CR7, which \citet{sobral15} have argued is young and metal-poor.

\subsection{Escape fraction}
In \citet{hu16}, we discussed what can be inferred about the escape
of ionizing photons if the ionized regions around the LAEs 
transition from optically thick to optically thin at a characteristic
Ly$\alpha$ luminosity. The intrinsic ionizing photon production
rate of a galaxy is related to its intrinsic Ly$\alpha$ luminosity through 
$\dot{N}_{\rm ion}\ ({\rm s}^{-1})=6.3\times 10^{12}\ L({\rm Ly}\alpha)\ ({\rm erg\ s}^{-1})$
\citep[e.g.,][]{matthee15}, assuming only that case~B is appropriate, that the IMF 
has a Salpeter power law at high mass, and that the Ly$\alpha$ luminosity is not
significantly reduced in the galaxy itself. If the last assumption is violated, 
then $\dot{N}_{\rm ion}$ will be higher. 

If the transition from optically thick to optically thin occurs at around
$\log L({\rm Ly\alpha}) =  43.5~{\rm erg~s}^{-1}$, as the data suggest,
then the ionizing photon rate is
$2\times 10^{56}~f_{\rm esc}\ ({\rm s}^{-1})$, where 
$f_{\rm esc}$ is the fraction of ionized photons that escape from the galaxy into 
the IGM\footnote{Note that there is a typographical error in \citet{hu16}, where 
this is given as $2\times 10^{46}~f_{\rm esc}\ ({\rm s}^{-1})$.
The remainder of their discussion is correct.}.
We expect that the ionized region generated from this will be large enough
that radiation damping wings from neutral gas in the IGM outside the ionized 
region may be neglected, unless the IGM is substantially neutral 
\citep{miralda98, haiman02}.
Thus, the crucial quantity is the scattering of the
Ly$\alpha$ profile by the residual neutral gas
in the ionized region \citep{haiman02, haiman05}.
Only the neutral gas at velocities lower than the galaxy
matters, and this scatters the blue side of the Ly$\alpha$ line.

The fraction of neutral hydrogen in the ionized zone is given by
\begin{equation}
x = 8.6\times10^{-4} C\,\left (r \over {\rm Mpc}\right )^{2} \,\left (\dot{N}_{ion}f_{esc}\over 10^{54}~{\rm s}^{-1}\right )^{-1}\ \,,
\end{equation}
where $C$ is the clumping factor, $r$ is the radial distance from the galaxy,
the recombination coefficient is computed at a temperature of $10^4$ K, and the
mean ionization cross section is that given by \citet{cen00}. This rises toward the 
outside edge of the zone and is smallest near the LAE itself. 

If we neglect density fluctuations of the gas inside the
ionized region and use the calculations of \cite{haiman02} corresponding to 
$\dot{N}_{\rm ion}f_{esc} = 1.4\times 10^{54}\  {\rm s}^{-1}$, then there is a very significant 
opacity of $\tau \sim 1000$ at the wavelength of the blue wing (solid line in
Figure~1 of \citet{haiman02}). However, density fluctuations within the ionized region reduce the 
average opacity, and a more realistic treatment of this effect reduces the value of  
$\tau$ to $\sim 2$ at the wavelength of the blue wing (long dashed line in Figure~1 of \citet{haiman02}). If we assume the transition from optically thick to optically thin ionized 
bubbles occurs at an optical scattering depth of $\tau = 1$, then we can scale the 
Haiman results to $\dot{N}_{\rm ion}f_{esc} = 3\times 10^{54}\ {\rm s}^{-1}$.
Combining this with our estimate of the ionizing photon rate at the transition,
we find $f_{esc} = 0.015$. However, the uncertainty in the calculation of the reduction
arising from density fluctuations is significant and should be borne in mind.

\section{Summary}
\label{secsum}
We have so far surveyed a $30~{\rm deg}^2$ area (out of $120~{\rm deg}^2$ planned) 
around the North Ecliptic Pole using the NB921 filter on HSC on the Subaru 
telescope as part of our HEROES program to search
for ultraluminous LAEs at $z=6.6$. Our follow-up observations of seven candidate LAEs
(out of 13) with the DEIMOS spectrograph on Keck~II showed that 
the spectral profiles of the five confirmed LAEs are wider than
those of lower luminosity LAEs at this redshift. We also found that one galaxy, 
NEPLA4, has a complex profile with an apparent blue wing that is similar to, but 
slightly wider than, the complex profile of the COLA1 galaxy found by \citet{hu16}
in the extended COSMOS field.

Combining our results from the NEP and COSMOS fields, we find that,
above a Ly$\alpha$ luminosity of $\log L$(Ly$\alpha)>43.5$~erg~s$^{-1}$, 
roughly a quarter (2 out of 8) of the LAEs have complex profiles with apparent
blue wings. This fraction is very similar to the fraction of LAEs with blue wings
at $z=0.3$, where IGM scattering is not significant, allowing 
the full profile emerging from the galaxy to be visible. The peak-to-peak velocity separations 
between the main and blue wing components are also similar to those seen in
the $z=0.3$ LAEs, suggesting that they may be good analogs
in terms of their inflow and outflow properties.

We can understand the existence of such complex Ly$\alpha$ profiles at $z=6.6$
if the ultraluminous LAEs are generating ionized zones around themselves  
with low enough neutrality that they do not significantly scatter the blue wings
\citep[e.g.,][]{haiman02,matthee15}. Then the full Ly$\alpha$ line would be able 
to emerge from the galaxy and be seen, even 
if the general IGM is substantially neutral \citep[e.g.,][]{cen00,haiman02}.

We also found that the Ly$\alpha$ profiles of the ultraluminous LAEs at $z=6.6$ 
are wider relative to those at lower luminosities at the same redshift,
suggesting that the more luminous LAEs have larger inflow and outflow
velocities than the lower luminosity LAEs. 

Using calculations by \citet{haiman02}, we matched the ionizing photon 
rate that corresponds to the Ly$\alpha$ luminosity at the
transition from optically thick to optically thin ionization bubbles with what would
be required to keep the optical scattering depth less than one, in order to estimate
the escape fraction of ionizing photons generated in the LAE.  We found a value
of $f_{esc}=0.015$. However, more sophisticated calculations of the density 
structure in the ionized regions are clearly needed.

The present samples are small, but spectroscopic follow-up of ultraluminous 
LAEs at $z=6.6$ selected from HSC observations is continuing.
In particular, the final HEROES sample should yield about 20 such galaxies
and about 7 additional blue-wing LAEs similar to NEPLA4 and COLA1.
Future modeling of the profiles should allow us to begin to determine the 
kinematic structure of galaxies at these redshifts.

\acknowledgements
We gratefully acknowledge support from NSF grants AST-1716093 
(E.~M.~H., A.~S.) and AST-1715145 (A.~J.~B) and from the Trustees
of the William F. Vilas Estate and the University of Wisconsin-Madison
Office of the Vice Chancellor for Research and Graduate Education
with funding from the Wisconsin Alumni Research Foundation (A.~J.~B.).
The NB921 filter was supported by KAKENHI (23244025) Grant-in-Aid for Scientific Research (A) through the Japan Society for the Promotion of Science (JSPS).  The NB816 filter was supported by a grant from Ehime University. 
We would like to thank Claudia Scarlata for permission to
quote unpublished data, David Sobral and Jorryt Matthee for 
extremely helpful comments on a first draft of the paper, and
the anonymous referee for a very useful and constructive
report which improved the paper.
The authors wish to recognize and acknowledge the very significant 
cultural role and reverence that the summit of Mauna Kea has always 
had within the indigenous Hawaiian community. We are most fortunate 
to have the opportunity to conduct observations from this mountain.

\end{document}